\documentclass[floatfix,twocolumn,showpacs,amsmath,amssymb,superscriptaddress,aps,prc]{revtex4-2}
\usepackage[dvips]{graphicx}
\usepackage{xcolor}
\usepackage{booktabs} 
\usepackage{longtable}
\usepackage{multirow}

\begin{document}
\title{A Quantization-Constrained Parameter Method for Studying $\alpha$ and Cluster Decay}
\author{Nguyen Thi Huyen Nga}
\author{Le Hoang Chien}\email{lhchien@hcmus.edu.vn}
\author{Nguyen Tri Toan Phuc}
\author{Chau Van Tao}
\affiliation{Department of Nuclear Physics, Faculty of Physics and Engineering Physics, University of Science, 700000 Ho Chi Minh City, Vietnam}
\affiliation{Vietnam National University, 700000 Ho Chi Minh City, Vietnam}
\date{\today}
\begin{abstract}
We propose a Quantization-Constrained Parameter Method (QCPM) for determining the Woods-Saxon (WS) potential depth $V_0$ and diffuseness $a$ in studies of $\alpha$ and cluster decay half-lives. In this approach, the parameters are derived analytically by imposing the Bohr--Sommerfeld (BS) quantization condition, thereby eliminating the need for parameter fitting. The calculated diffuseness values exhibit a strong dependence on nuclear shell structure. The reliability of the QCPM-derived WS potential is validated through optical model analyses of elastic $\alpha$ scattering data. The resulting $\alpha$ and cluster decay half-lives show good agreement with experimental values. Our results highlight the significant role of daughter-nucleus deformation in improving the consistency between theoretical predictions and experimental data. In addition, we introduce a modified Woods--Saxon (mWS) potential that effectively approximates the surface behavior of folding potentials incorporating the nuclear medium effect. This modification leads to closer agreement with the measured half-lives. Free of adjustable parameters such as $V_0$ and $a$, the QCPM enhances the predictive power of phenomenological potentials for $\alpha$ and cluster decay studies.
\end{abstract}
\maketitle

\section{Introduction}
\label{sec1}

$\alpha$ and cluster radioactivity are among the dominant decay modes of heavy and superheavy nuclei. These processes provide valuable insights into the nuclear structure of parent nuclei and the dynamics of quantum tunneling. In addition, these decay modes are essential tools for the identification of new elements \cite{Smits24,Giu19,Sam21,Zhang21}. Consequently, the $\alpha$ and cluster emissions remain important topics, attracting considerable attention from both experimental and theoretical studies \cite{Yang24,Belli25,Zhang25,Wang24,Ismail24,Seif25}.

Generally, $\alpha$ and cluster decays are described as quantum tunneling processes of preformed $\alpha$ particles or heavier clusters in the nuclear surface region \cite{Gamow28,Gurney28}. Because of the complexity of the nuclear many-body problem, most theoretical approaches reduce the system to two-body calculations, where the cluster interacts with the daughter nucleus \cite{Buck89,Buck92,Mohr06,Ni09,Dens15,Poenaru18,Delion18}. Consequently, the choice of the nuclear potential between the cluster and the residual nucleus becomes a critical factor in improving the accuracy of theoretical half-life predictions. 

In addition to double-folding potentials \cite{Mohr00,Mohr06,Khoa01}, phenomenological nuclear potentials, commonly expressed using three-parameter Woods--Saxon (WS) or cosh-parameterized functions, are widely adopted in the literature \cite{Buck92,Ni09,Coba12,Dens15,Sun16}. Most studies involve adjusting the potential depth and surface diffuseness parameters to reproduce experimental half-lives. While this fitting approach often yields good agreement with existing data, it may compromise the model's predictive power, particularly for cluster decay, where experimental data are usually limited or unavailable.

The surface diffuseness plays a significant role in both phenomenological and microscopic nuclear models \cite{Adam14,Dahm17,Alad19}. In most studies, it is treated as a fixed parameter, with values ranging from 0.4 to 0.8 fm. Specifically, the WS nuclear potentials with the fixed diffuseness $a=0.4929$ fm \cite{Dens05}, $a=0.6$ fm \cite{Qian11,Ni09}, and $a=0.7$ fm \cite{Coba12} have been used. In Ref.~\cite{Buck95}, the authors used a modified WS potential with a fixed value $a=0.65$ fm. Other works also performed calculations using the cosh-parameterized potential with constant diffuseness $a=0.4$ fm \cite{Buck93}, $a=0.5958$ fm \cite{Sun16}. Beyond fixed values, alternative parameterizations have been proposed to include isospin or mass dependence to the diffuseness \cite{Ni10,Dens25}, which aim to account for structural variations. These diverse choices emphasize the sensitivity of calculated decay half-lives to surface properties and point to the need for a more systematic treatment. Rather than being treated as a constant, surface diffuseness should reflect the underlying nuclear structure of emitters.

In this work, we propose the Quantization-Constrained Parameter Method (QCPM) to consistently determine the potential depth and diffuseness. This approach is based on the requirement that the nuclear potential must satisfy the Bohr--Sommerfeld (BS) quantization condition, while simultaneously establishing an analytical relationship between the potential depth and diffuseness. By avoiding parameter fitting, the method reduces model dependence and enhances the predictive power of phenomenological potentials. The QCPM is thus expected to strengthen the connection between phenomenological potentials and the underlying nuclear structure.

Additionally, we take into account the deformation of daughter nuclei through angle-dependent modifications to both the nuclear and Coulomb potentials. The decay width is then obtained through integration over all possible orientations, yielding an orientation-averaged value. To evaluate the effect of nuclear deformation, the results are systematically compared with those calculated under the assumption of spherical nuclei.

The paper is organized as follows. Sec. \ref{sec2} outlines the theoretical framework, focusing on the quantization-based derivation of the potential depth and diffuseness and a brief overview of the semi-classical cluster decay model. Sec. \ref{sec3} presents the calculated potential parameters and the analysis of $\alpha$ and cluster decay half-lives. Finally, the conclusions are summarized in Sec.~\ref{sec4}.

\section{Methods}
\label{sec2}

\subsection{A Quantization-Constrained Parameter Method}
The effective potential \(V_{\rm eff}(r)\) between cluster and daughter nuclei is given by the sum of the nuclear, Coulomb, and centrifugal potentials
\begin{align}
	V_{\rm eff}(r) = V_N(r) + V_C(r) + \frac{\hbar^2}{2\mu} \left( \ell + \frac{1}{2} \right)^2 \dfrac{1}{r^2}.\label{eq1}
\end{align}
Here, the centrifugal term incorporates the Langer correction, replacing $\ell(\ell+1)/r^2$ with $(\ell+1/2)^2/r^2$, which improves agreement between the Wentzel–Kramers–Brillouin (WKB) approximation and exact solutions \cite{Langer37}. $\ell$ denotes the orbital angular momentum. For clusters with spin \(I_c=0\) and positive parity (\(\pi_c=+\)), the allowed values of $\ell$ are determined by the selection rules
\(
|I_p - I_d| \le \ell \le I_p + I_d \mathrm{~and~} \pi_p = \pi_d (-1)^\ell
\), where \(I\) and \(\pi\) correspond to the spin and parity of the parent (\(p\)) and the daughter (\(d\)) nuclei, respectively. \( \mu = {M_c M_d}/{(M_c+M_d}) \) is the reduced mass of the cluster--daughter system, with \(M_c\) and \(M_d\) denoting the masses of the cluster and the daughter nucleus, respectively. The Coulomb potential \(V_C(r)\) is calculated using the standard expression for a uniformly charged sphere with radius \(R_{0} \) \cite{Glendenning04,Buck95,Mohr06,Ni09}, given by
\begin{align}
V_C(r) &=
	\dfrac{Z_c Z_d e^2}{r}, \qquad \qquad \qquad~  r \geq R_{0},  \notag \\[9pt]
	&= \dfrac{Z_c Z_d e^2}{2R_{0}} \left( 3 - \frac{r^2}{R_{0}^2} \right), \quad r < R_{0},\label{eq2}
\end{align}
where \(Z_c\) and \(Z_d\) correspond to the atomic numbers of the cluster and daughter nuclei. The nuclear potential \( V_N(r)\) is commonly described by the standard WS form
\begin{equation}
	V_N(r) = \dfrac{-V_0} {1+ \exp\!{\left [(r - R_0)/a \right ] }}.\label{eq3}
\end{equation}
The radius parameter \(R_0\) is defined as 
\begin{equation}
	R_0 = R_d+\Delta R,\label{eq4}
\end{equation}
with
\begin{equation}
\Delta R= R_p - R_d.\label{eq5}
\end{equation}
Here, \(R_p\) and \(R_d\) denote the \textcolor{black}{spherical} radii of the parent and daughter nuclei, respectively. \textcolor{black}{From Eqs.~\eqref{eq4}–\eqref{eq5}, one sees that \(R_0\) reduces to \(R_p\) when deformation effects of the daughter nucleus are neglected}. The radii are parameterized as follows \cite{Wang13}  
\begin{align}
	R_i =& ~ 1.226 A_i^{1/3}+2.86 A_i^{-2/3} 
	+0.99 \Delta E_i /A_i \notag \\[5pt] 
	&-1.09 \dfrac{(A_i-2 Z_i) 2Z_i}{A_i^2}  \quad  \text{with~} i=p,d, \label{eq6}
\end{align}
where \(A_{p(d)}\) is the mass number of the parent (daughter) nucleus. The shell correction energies $\Delta E_{p(d)}$ are taken from Refs.~\cite{Wang2010,WS3}. The expression in Eq.~\eqref{eq6} incorporates shell and isospin effects, thereby going beyond the simple $r_0 A^{1/3}$ form commonly used in the WS potential. Such a parameterization has been shown to provide a more accurate reproduction of experimental data than those obtained from the microscopic nuclear mass model or the relativistic mean-field approach \cite{Wang13}. The definition in Eq.~\eqref{eq4} enables the inclusion of daughter-nucleus deformation through the multipole expansion of \(R_d\) and accounts for the finite size of the emitted cluster via \(\Delta \! R\).
   
Since \(V_0\) and \(a\) are more uncertain and typically need to be adjusted to best fit experimental data, we propose the QCPM to determine these parameters. Specifically, \textcolor{black}{for distances \(r \geq R_0\)}, by substituting \(V_N(r)\) from Eq.~\eqref{eq3} into Eq.~\eqref{eq1} and taking the first derivative of \( V_{\rm eff} \) with respect to $r$, we obtain
\begin{align}
	\dfrac{d V_{\rm eff}}{dr} = \frac{V_0}{a} \dfrac{e^{(r - R_0)/a}}
	{\big [ 1+e^{(r - R_0)/a} \big ]^2
	} 
	- \frac{Z}{r^2} - \frac{2 C_{\ell} }{r^3},\label{eq7}
\end{align}
where $Z = Z_c Z_d e^2$ and $C_{\ell} = \dfrac{\hbar^2}{2\mu} \left( \ell + \dfrac{1}{2} \right)^2$.
At large distances near the Coulomb barrier position  $r \sim R_B$, the last term in Eq.~\eqref{eq7} contributes minimally and can therefore be neglected. Accordingly, by solving the equation \( \dfrac{d V_{\rm eff}}{dr} \Big|_{r=R_B} = 0 \) for \( V_0 \), we obtain
\begin{equation}
	V_0 = \dfrac{Z a}{R_B^2} 
	\dfrac{\Big [~ 1+\exp \! {\left [(R_B - R_0)/a \right ]}~ \Big ]^2}
	{\exp \! {\left [(R_B - R_0)/a \right ] }}.\label{eq8}
\end{equation}

The next step is to determine a relationship between the barrier position \(R_B\) and the surface diffuseness \(a\). By inserting Eq.~\eqref{eq3} incorporating Eq.~\eqref{eq8} into Eq.~\eqref{eq1}, the Coulomb barrier height \( V_B \) can be obtained as follows
\begin{align}
	V_B = V_{\rm eff}(r) \Big |_{r=R_B}
	=&-
	\dfrac{Z a}{R_B^2} \Bigg \{ \dfrac{1+\exp\!{\left [(R_B - R_0)/a \right] }
	}{\exp\!{\left [(R_B - R_0)/a \right]}} \Bigg \} \nonumber
	\\ &+ \frac{Z}{R_B} + \frac{C_{\ell}}{R_B^2}.\label{eq9}
\end{align}
At the barrier position \(R_B\), the exponential term \( \exp\!{\left [(R_B - R_0)/a \right ]} \) is much greater than 1. Therefore the ratio 
\begin{align}
\dfrac{1+\exp\!{\left [(R_B - R_0)/a \right] }
}{\exp\!{\left [(R_B - R_0)/a \right]}} \simeq 1.\label{eq10}
\end{align}
By rearranging Eq.~\eqref{eq9} using Eq.~\eqref{eq10}, one can obtain the quadratic equation for \(R_B\) as
\begin{align}
V_B R_B^2 - Z R_B + (Z a - C_{\ell}) = 0,\label{eq11}
\end{align}
which leads to the \textcolor{black}{positive} solution 
\begin{align}
	R_B = \dfrac{1}{2}
	\bigg [
	\dfrac{Z}{V_B} + \sqrt{
		\Big ( \dfrac{Z}{V_B} \Big )^2
		-4 \dfrac{( Za-C_{\ell} )}{V_B}
	}~
	\bigg ].\label{eq12}
\end{align}

A similar expression to Eq.~\eqref{eq12} can be found in Ref.~\cite{Rowley15}, where it is used in the study of low-energy fusion reactions. This supports the reliability of Eq.~\eqref{eq10}.
A combination of Eqs.~\eqref{eq8} and \eqref{eq12} shows that the nuclear potential depth \(V_0\) can be determined once the Coulomb barrier height \(V_B\) and the surface diffuseness \(a\) are specified. The Coulomb barrier height can typically be extracted with high accuracy from elastic scattering and fusion data or from folding potential calculations. Consequently, the nuclear potential in Eq.~\eqref{eq3} is fully determined if the diffuseness \(a\) is known. 

In semi-classical studies of the quasi-bound state of a cluster--daughter system, the nuclear potential is constrained to reproduce the quasi-bound wave function associated with an experimental decay energy $Q$.	Within the potential pocket, the wave function must form a standing wave with a quantized number of nodes, as required by the well-known BS quantization condition \cite{Buck95}. Physically, the BS condition establishes a fundamental link between the quantized nature of the quasi-bound state of the preformed cluster and the underlying nuclear potential.
	
In this work, the BS condition is employed to determine the diffuseness parameter $a$ entirely from quantum-mechanical principles, without the need to fit to data, as follows
 \begin{align}
 	& \int_{r_1}^{r_2} \sqrt{\frac{2\mu}{\hbar^2}} 
 		\bigg [ Q 
 		+ \dfrac{V_0} {1+ \exp \! \big [ (r - R_0)/a \big ] }
 	 - V_C(r) \notag \\[9pt]
 	 & - \dfrac{\hbar^2}{2\mu} \left( \ell + \frac{1}{2} \right)^2 \dfrac{1}{r^2}
 	\bigg ]^{1/2} dr 
 	 = (G-\ell+1) \frac{\pi}{2},\label{eq13}
 \end{align}
\textcolor{black}{where the positions $r_1$ and $r_2$ denote the first (inner) and second classical turning points satisfying $V_\text{eff}(r)=Q$, with $Q$ being the decay energy. A third turning point $r_3$, corresponding to the outer solution of the same condition, enters in Eq.~\eqref{eq17}}. The global quantum number \(G\) arises from the Pauli exclusion principle, ensuring that the nucleons in the cluster occupy orbitals above those of the core nucleons \cite{Wildermuth1977,Mohr06,Buck93}, given as: 
 \begin{equation}
	G = 2n + \ell = \sum_{i=1}^{A_c} \left( g_i^{(A_d + A_c)} - g_i^{(A_c)} \right). \label{eq14}
\end{equation}
Here, $A_d$ and $A_c$ stand for the nucleon numbers of the core and cluster nuclei, respectively. 
\( g_i^{(A_d + A_c)} \) is the oscillator quantum number of a cluster nucleon orbiting the core. The values of $g_i^{(A_d + A_c)}$ are determined from the neutron and proton shell structure of the parent nucleus. Specifically, we take $g_i^{(A_d + A_c)}=4$ for nucleons in the 50 $\leq N_p, Z_p \leq$ 82 shell, $g_i^{(A_d + A_c)}=5$ for nucleons in the 82 $< N_p, Z_p \leq$ 126 shell, and $g_i^{(A_d + A_c)}=6$ for nucleons in the $N_p >$ 126 shell. 

In Eq.~\eqref{eq14}, \( g_i^{(A_c)} \) denotes the internal quantum number of a cluster nucleon, determined according to the shell structure of the cluster. For instance, in the $^{14}$C cluster decay of $^{222}$Ra, two protons occupy the $g_i^{(A_c)}=0$ shell $(0s_{1/2})$ while the remaining four protons are in the $g_i^{(A_c)}=1$ shell $(0p_{3/2})$. Among the eight neutrons in the $^{14}$C cluster, the first six follow the same configuration as the six protons, and the remaining two neutrons occupy the $0p_{1/2}$ orbital, corresponding to $g_i^{(A_c)}=1$. As a result, the total internal quantum number is $\sum_{i=1}^{14} g_i^{(A_c)} = 10$. A simpler example is the $\alpha$ decay. Two protons and two neutrons of the $\alpha$ cluster occupy the $0s_{1/2}$ orbitals, which leads to $\sum_{i=1}^{4} g_i^{(A_c)} = 0$.

We recall that \(V_0\) in Eq. \eqref{eq13} is expressed as a function of the Coulomb barrier height \(V_B\) and the surface diffuseness \(a\), as given in Eqs.~\eqref{eq8} and \eqref{eq12}. This relationship imposes a quantization condition on the diffuseness \(a\) within the semi-classical model. 
Although the QCPM is derived here for the WS potential, the approach can be generalized to other phenomenological forms. In such cases, the definition of $V_0$ in Eq.~\eqref{eq8} must be modified accordingly, as illustrated later.

\subsection{Cluster decay half-life}
Within the framework of the semi-classical cluster model \cite{Gurvitz87,Buck92}, the decay half-life for spherical daughter and cluster nuclei is related to the penetration probability \( P \) by
 \begin{equation} T_{1/2} = \frac{ \hbar \text{ln} 2}{S_c F \big [\hbar^2/(4\mu) \big ] P}. \label{eq15}
\end{equation} 
Here, \(S_c\) is the cluster preformation factor while the normalization factor \(F\) reads
\begin{equation} F = \left[ \int_{r_1}^{r_2} \dfrac{dr}{2 \sqrt{ 2\mu/\hbar^2 \big ( Q-V_{\rm eff}(r)\big ) }} \right]^{-1}. \label{eq16} \end{equation} 
The penetration probability $P$ is computed using the standard WKB approximation, given by
\begin{equation}
	P = \exp\left(-2 \int_{r_2}^{r_3} \sqrt{\frac{2\mu}{\hbar^2} \big (V_{\rm eff}(r) - Q \big )} \, dr \right).\label{eq17}
\end{equation}
\textcolor{black}{We note that the same turning points $r_1$, $r_2$, and $r_3$ also enter in Eqs.~\eqref{eq16} and \eqref{eq17}}. For further details on these formulas, readers may refer to Refs.~\cite{Buck92,Chien22}.

For deformed nuclei, deformation effects can be included by modeling the nuclear and Coulomb potentials as angle-dependent functions. Assuming the cluster is spherical and the daughter nucleus has an axially symmetric shape, the radius parameter \(R_0\) in Eq. \eqref{eq4} is now modified as
\begin{equation}
R(\theta) = R_d \left [  1+ \sum_{\lambda=2,4} {\beta_\lambda Y_{\lambda 0}(\theta)} \right] + \Delta R. \label{eq18}
\end{equation}
The Coulomb potential for quadrupole- and hexadecapole-deformed daughter nuclei can be accordingly expressed to lowest order in $\beta_2$ and $\beta_4$ \cite{Dens25} as
\begin{align}
	V_C(r, \theta) &= \frac{Z}{r} \Bigg [ 1 + \frac{3R_0^2}{5r^2} \beta_2 Y_{20}(\theta)  + \frac{3R_0^4}{9r^4} \beta_4 Y_{40}(\theta) \Bigg ],
	\notag \\[5pt] 
	& \qquad \qquad \qquad \qquad \qquad \quad r \geq R(\theta), \label{eq19}
\end{align}
and	
\begin{align}
	&V_C(r, \theta) \simeq \frac{Z}{R(\theta)} 
	\left [ \frac{3}{2} - \frac{r^2}{2R^2(\theta)}   +~ \frac{3R_0^2}{5R^2(\theta)} \beta_2 Y_{20}(\theta)  
	  \right.
	\notag \\[5pt]
	& \left. 
\times	\left( 2 - \frac{r^3}{R^3(\theta)} \right)
	+~ \frac{3R_0^4}{9R^4(\theta)} \beta_4 Y_{40}(\theta)  
	\left( \frac{7}{2} - \frac{5r^2}{2R^2(\theta)} \right) \right],
	\notag \\[5pt]
	& \qquad \qquad \qquad \qquad \qquad \qquad \qquad \quad r < R(\theta), \label{eq20}
\end{align} 
where \(R_0\) and \(R(\theta)\) are obtained from Eqs. \eqref{eq4} and \eqref{eq18}, respectively. In Eqs.~\eqref{eq18}, \eqref{eq19}, and \eqref{eq20}, $\theta$ represents the angle between the symmetry axis and the line connecting the cluster and daughter nuclei. Taking deformation effects into account, we apply Eq.~\eqref{eq15} to calculate the cluster decay half-life. However, the normalization factor $F$ in Eq.~\eqref{eq16} and the penetration probability $P$ in Eq.~\eqref{eq17} must be modified accordingly. Specifically, these quantities, $F$ and $P$, are evaluated for each orientation angle $\theta$, and the final results are obtained by averaging over the range $ 0 \leq \theta \leq \pi$ \cite{Dens05,Coba12}, as follows
\begin{equation} \overline{F} = \dfrac{1}{2}\int_{0}^{\pi} F(\theta) \sin(\theta)  d\theta,~ 
	\overline{P} = \dfrac{1}{2}\int_{0}^{\pi} P(\theta) \sin(\theta)  d\theta.
	 \label{eq21} \end{equation} 
 
\section{Results and Discussion}
\label{sec3}
\subsection{The constrained potential depth and diffuseness}
\label{sec3a}
\begin{figure}[htb]
	\centering
	\includegraphics[width=8.9cm]{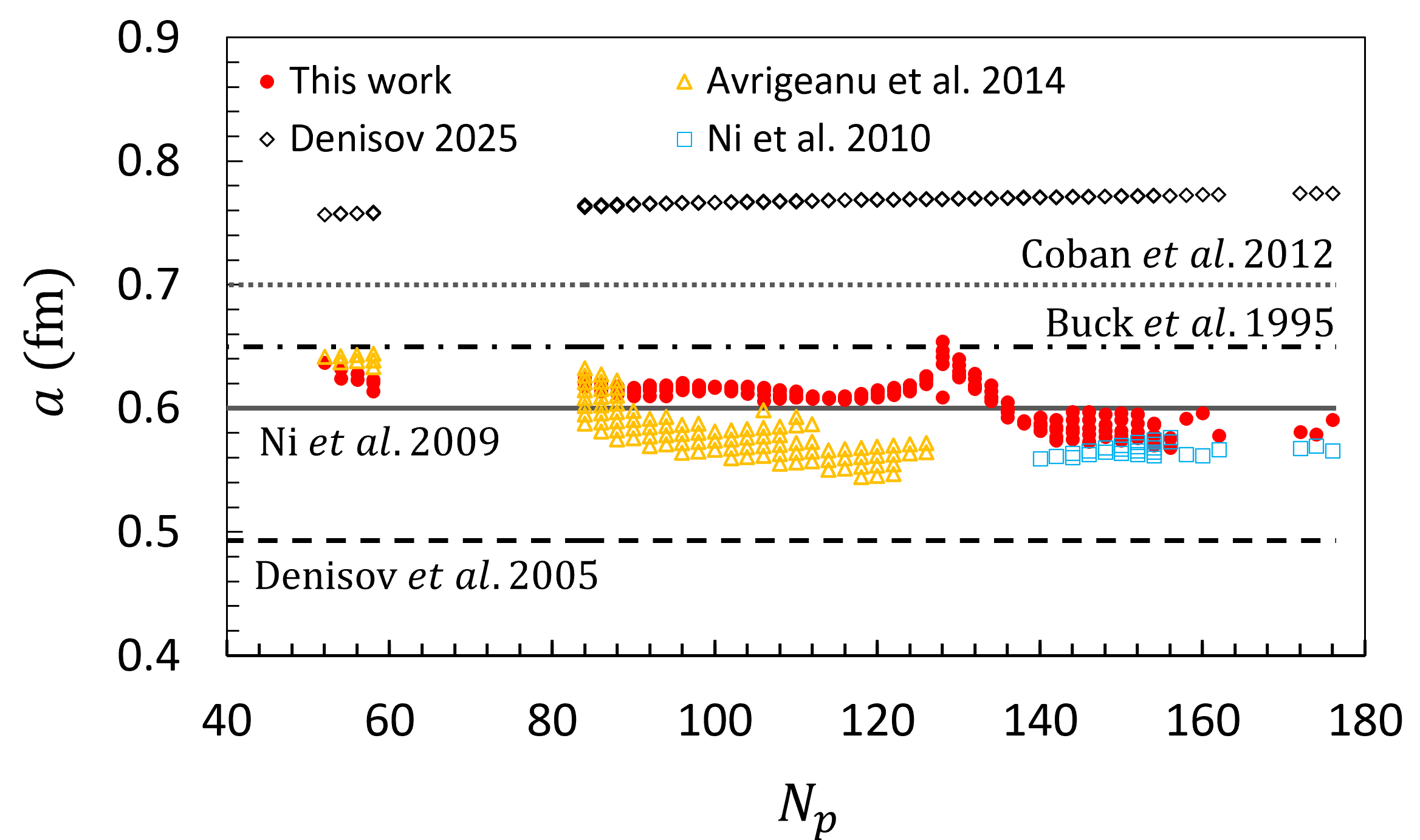}
	\caption{Constrained diffuseness values (red solid points), calculated using Eqs.~\eqref{eq8}, \eqref{eq12}, and \eqref{eq13}, are compared with various values in the literature: solid line \cite{Ni09}, dotted line \cite{Coba12}, dashed line \cite{Dens05}, dash-dotted line \cite{Buck95}, open squares \cite{Ni10}, open diamonds \cite{Dens25}, and open triangles \cite{Avri14}.}
	\label{fig1_a}
\end{figure}

First, we apply the QCPM to calculate the diffuseness $a$ for the $\alpha$--daughter WS potential, using Eqs.~\eqref{eq8}, \eqref{eq12}, and \eqref{eq13}. In calculations, the decay energy $Q$ is taken from Ref.~\cite{Wang21}. \textcolor{black}{The Coulomb barrier height $V_B$ in Eq.~\eqref{eq12} is determined from the global $\alpha$--nucleus potential with parameters listed in Table~II of Ref.~\cite{Avri14}. Since this potential is energy dependent, $V_B$ is evaluated using $Q$ as an input}. A wide range of even--even parent nuclei with mass numbers $100 \leq A_p \leq 290$ ($50 \leq N_p \leq 180$) is considered. The calculated diffuseness values are compared with various values in the literature, as illustrated in Fig.~\ref{fig1_a}. 

The constrained diffuseness values in this study range from 0.57 to 0.65 fm, generally falling within the range of previously reported values. Interestingly, the calculated diffuseness vary around the best-fit values $a=0.5958$ fm \cite{Sun16} and $a=0.65$ fm \cite{Buck95} (dash-dotted line). Furthermore, our results are close to those obtained from parameterizations in Refs.~\cite{Avri14,Ni10} (open triangles and squares).

As shown in Fig.~\ref{fig1_a}, the calculated diffuseness values vary with the shell structure of the $\alpha$ emitters. The values fluctuate with the neutron number and increase sharply near the $N_p = 126$ shell closure, which appears linked to shell effects. This observation is consistent with a previous study that demonstrates the shell-dependent variation in the depth of a square-well potential \cite{Tian24}. Consequently, these results suggest that shell effects play a significant role in determining the $\alpha$--nucleus potential. 

\begin{figure}[htb]
	\centering
	\includegraphics[width=8.8cm]{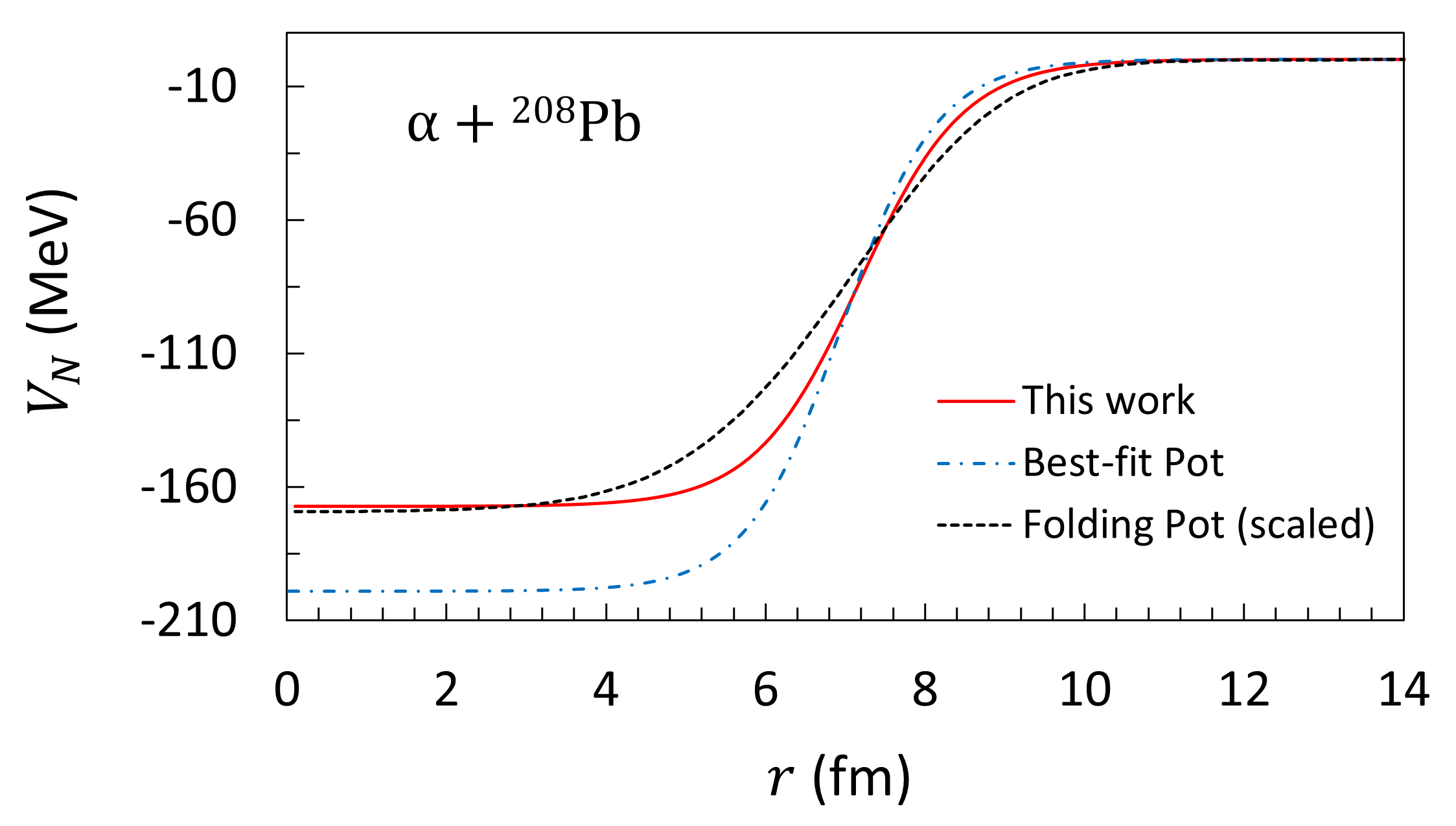}
	\caption{Comparison of the $\alpha+^{208}$Pb nuclear potential models. The red solid line shows the WS potential in Eq.~\eqref{eq3} using the $V_0$ and $a$ values derived from the QCPM. The black dotted line represents the folding potential scaled by a factor $\lambda=0.925$ due to application of the BS quantization condition \cite{Chien22}. The blue dash-dotted line stands for the potential with parameters fitted to experimental $\alpha$ decay half-lives \cite{Sun16}.}
	\label{fig2_p}
\end{figure}

In Fig. \ref{fig2_p}, the $\alpha+^{208}$Pb nuclear potential with the depth $V_0$ and diffuseness $a$ obtained from the QCPM (red solid line) is compared with the standard folding potential \cite{Chien22} (black dotted line) and the best-fit potential \cite{Sun16} (blue dash-dotted line). In the present model, $V_0$ and $a$ are determined consistently using Eqs.~\eqref{eq8}, \eqref{eq12}, and \eqref{eq13}. \textcolor{black}{The folding potential is calculated using the density- and energy-dependent CDM3Y3 interaction with a finite-range exchange term \cite{Khoa97}, following Eqs.~(10)--(15) in Ref.~\cite{Chien22}. To satisfy the BS condition, the resulting potential is further scaled by 0.925, as illustrated by the red dotted line in Fig.~1 of Ref.~\cite{Chien22}. The best-fit potential is based on the cosh-parameterized form given in Eqs.~(6), (8), and (11) of Ref.~\cite{Sun16}, with parameters adjusted to reproduce the experimental $\alpha$ decay half-lives.}

We recall that $\alpha$ decay half-lives of heavy nuclei are mainly determined by the nuclear potential in the surface region $r \geq 7$ fm. At such distances, our potential lies within the range defined by the best-fit and folding potentials. Therefore, the QCPM-derived potential is expected to be sufficiently reliable for $\alpha$ decay calculations.

To further compare these potential models, we also consider the volume integral \cite{Mohr00}. This quantity is independent of specific potential parameterizations and is defined as follows
\begin{equation}
	J_R = \frac{4\pi}{A_c A_d} \int_{0}^{\infty} V_N(r) r^2 \, dr. \label{eq22}
\end{equation}
The volume integral of our potential model for the $\alpha+^{208}$Pb system is $J_{R} = -336.1~\text{MeV}\,\text{fm}^3$. This value is comparable to that of the scaled folding potential ($J_{R} = -329.7~\text{MeV}\,\text{fm}^3$) \cite{Chien22}. Compared with the best-fit potential \cite{Sun16} with $J_{R} = -362.1~\text{MeV}\,\text{fm}^3$, our $J_{R}$ is roughly 7$\%$ smaller. We recall that the best-fit potential was optimized for the best agreement with $\alpha$ decay half-life data but it did not incorporate the BS quantization condition. This omission may have contributed to the larger volume integral compared with our case.

\begin{table}[ht]
	\caption{The best-fit parameters of the imaginary part used in the optical potential analysis are presented. $E_{\rm lab}, V_0$, and $W_0$ are given in MeV while $R_{\rm 0}$, $R_{\rm W}$, $a$, and $a_{\rm W}$ are in fm. The volume integrals $J_R$ and $J_I$ are calculated in units of $\text{MeV}\, \text{fm}^3$. 
		\label{tb1}}
	\begin{ruledtabular}
		\begin{tabular}{cccccccccc}
			$E_{\rm lab}$ & $V_0$\footnotemark[1] & $R_{\rm 0}$ & $a$\footnotemark[1] & $-J_{\rm R}$
			& $W_0$ & $R_{\rm W}$ & $a_{\rm W}$ & $-J_{\rm I}$
			\vspace{0.05 in} \\ 
			\hline \hline \\
			\multicolumn{1}{c}{$\alpha+^{140}{\!\rm Ce}$}\\
			$19$ & $151.06$ & $6.37$ & $0.619$ & $319.5$ & $11.58$ & $7.97$ & $0.80$ & $48.3$ \\
			&&&&328.1\footnotemark[2]&&&&43.1\footnotemark[2]& \\
			&&&&383.3\footnotemark[4]&&&&59.3\footnotemark[4]&	
			\vspace{0.05 in} \\
			\multicolumn{1}{c}{$\alpha+^{144}{\!\rm Sm}$}\\
			$19.87$ & $148.50$ & $6.46$ & $0.618$ & $317.7$ & $11.97$ & $8.44$ & $0.81$ & $57.1$  \\
			&&&&337.3\footnotemark[3]&&&&56.4\footnotemark[3]& \\
			&&&&331.4\footnotemark[3]&&&&58.1\footnotemark[3]&	\\
			&&&&330.8\footnotemark[3]&&&&58.3\footnotemark[3]& \\
			&&&&384.4\footnotemark[4]&&&&58.7\footnotemark[4]&		
			\vspace{0.05 in}\\
			\multicolumn{1}{c}{$\alpha+^{208}{\!\rm Pb}$}\\			
			23.6 & $167.19$ & $7.17$ & $0.654$ & $336.1$ & $12.07$ & $9.20$ & $0.81$ & $51.0$ \\
			&&&&383.6\footnotemark[4]&&&&52.7\footnotemark[4]&					
			\vspace{0.05 in}\\
			27.6 &&&&&$12.87$&$9.21$&$0.81$&$54.5$ \\	
			&&&&380.2\footnotemark[4]&&&&56.1\footnotemark[4]&	 \\	
		\end{tabular}
	\end{ruledtabular}
	\footnotetext[1]{Calculated by Eqs.~\eqref{eq8}, \eqref{eq12}, and \eqref{eq13}}
	\footnotetext[2]{\cite{Mohr13b}, $^{\text{c}}$ \cite{Kiss22}, $^{\text{d}}$ \cite{Avri14}; obtained from scattering analysis}
\end{table}
Next, we test the validity of the constrained diffuseness based on an optical potential analysis. Specifically, the potential derived in this study is used to describe the real part of the optical potential, while the imaginary part, which accounts for all non-elastic channels, is modeled using the standard WS form. The optical potential is thus expressed as follows
\begin{equation}
	U(r) = -\frac{V_0}{1 + \exp\!\left(\frac{r - R_0}{a}\right)} 
	- i\frac{W_0}{1 + \exp\!\left(\frac{r - R_W}{a_W}\right)}. \label{eq23}
\end{equation}
In this single-channel calculation, the two parameters $V_0$ and $a$ of the real part are not adjustable but are instead determined using Eqs.~\eqref{eq8}, \eqref{eq12}, and \eqref{eq13}. Only three parameters $\{W_0, R_W, a_W\}$ of the imaginary part are varied to obtain the best fit to scattering data. Several sets of $\alpha$--nucleus elastic scattering data at energies near the Coulomb barrier are considered for this analysis. We note that the real part of the optical potential derived from the QCPM is applicable only to scattering systems possessing the corresponding quasi-bound states. Therefore, the $\alpha+^{140}\!\mathrm{Ce}$, $^{144}\mathrm{Sm}$, and $^{208}\mathrm{Pb}$ systems are adopted for this study \cite{Gua94,Kiss22,Karcz72}. \textcolor{black}{These nuclei are known to be spherical \cite{Moller16}, which makes them well suited for testing the QCPM-derived diffuseness within the single-channel calculation.} All optical model calculations are performed using the code ECIS97 written by Raynal \cite{Raynal72}.

The best-fit imaginary parameters are listed in Table~\ref{tb1}. To facilitate comparison with other potentials in the literature, we calculate and discuss the volume integrals $J_R$ and $J_I$. For the imaginary parts, our $J_I$ values are in good agreement with those from previous studies. This indicates the imaginary potentials used in this work are consistent with other findings \cite{Kiss22,Mohr13b,Avri14}.

Regarding the real part of the potential, the calculated volume integrals $J_R$ are systematically smaller than those listed in Table~\ref{tb1}. This is because the potential parameters are not freely fitted but are constrained by Eqs.~\eqref{eq8}, \eqref{eq12}, and \eqref{eq13}, which are based on the BS condition associated with a quasi-bound state treatment. Therefore, consistency with $J_R$ values of potentials derived from scattering analyses lies beyond the scope of this work. Nevertheless, this discrepancy is expected not to noticeably affect the reliability of the WS potential obtained from the QCPM in describing \(\alpha\) decay. 
\begin{figure}[htb]
	\centering
	\includegraphics[width=8.8cm]{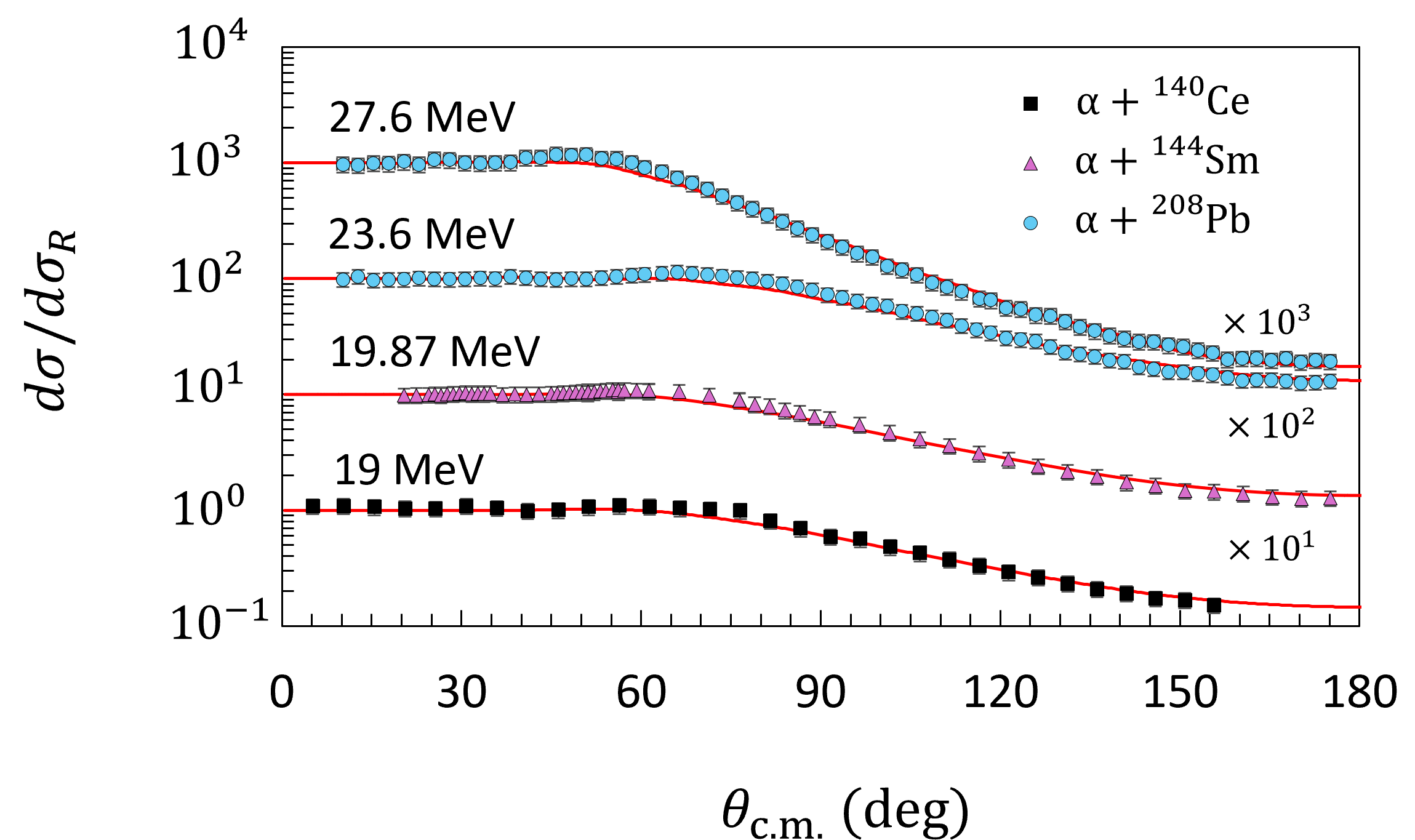}
	\caption{Optical model description of the elastic $\alpha+^{140}$Ce, $^{144}$Sm, and $^{208}$Pb scattering cross sections at energies near the Coulomb barrier. $d \sigma / d \sigma_R$ represents the differential cross section relative to the Rutherford cross section. Experimental data are taken from Refs.~\cite{Gua94,Kiss22,Karcz72}.}
	\label{fig3_e}
\end{figure}

In Fig. \ref{fig3_e}, the results obtained using the optical potential are compared with the experimental elastic scattering cross section data for $\alpha$ particles incident on the $^{140}$Ce, $^{144}$Sm, and $^{208}$Pb nuclei \cite{Gua94,Kiss22,Karcz72}. The calculations performed with our potential model show good agreement with the measured data across a wide range of scattering angles. This supports the reliability of the real part of the optical potential in describing the $\alpha$--nucleus interaction at low energies. Consequently, the diffuseness obtained from the QCPM can be considered sufficiently accurate.

As shown previously in Fig.~\ref{fig1_a}, the diffuseness exhibits a clear dependence on shell structure. This indicates that different choices of the parameter can considerably affect the reliability of $\alpha$ decay half-life calculations, particularly near the $N_p=126$ shell closure. To quantify this impact, we determine the logarithmic ratio between calculated (cal) and experimental (exp) $\alpha$ decay half-lives ($\Delta {\rm log}_{10}(T_{1/2}) = {\rm log_{10}}(T_{1/2}^{\rm cal}/T_{1/2}^{\rm exp})$) for various values of the diffuseness $a$. 
\begin{figure}[htb]
	\centering
	\includegraphics[width=8.8cm]{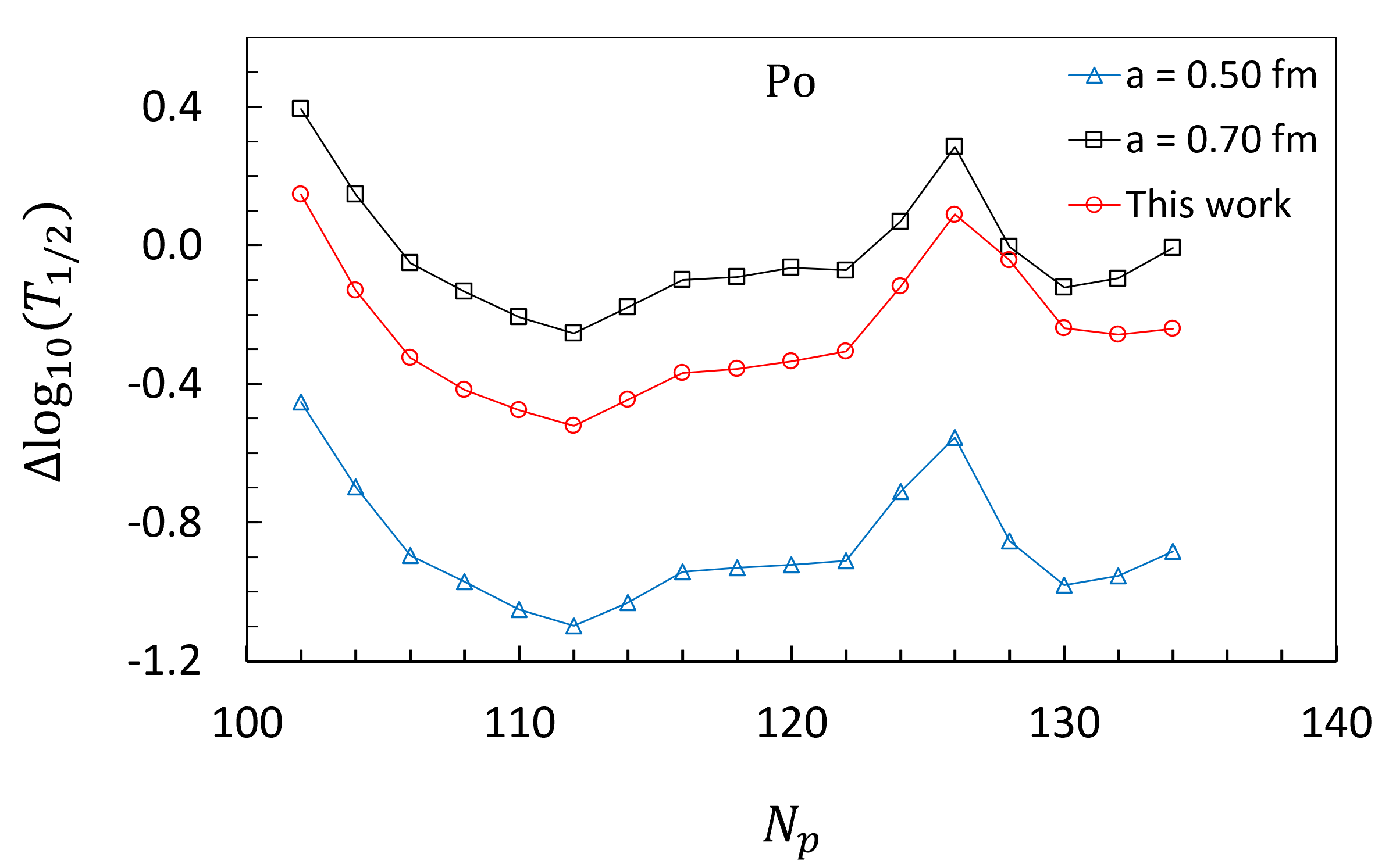}
	\caption{Logarithmic ratios of the calculated $\alpha$ decay half-lives to the experimental data for even--even Po isotopes. The ratio is defined as $\Delta {\rm log}_{10}(T_{1/2}) = {\rm log_{10}}(T_{1/2}^{\rm cal}/T_{1/2}^{\rm exp})$. Experimental data are taken from Ref.~\cite{Kondev21}.}
	\label{fig4_dt}
\end{figure}

In this comparison, two fixed diffuseness values, $a=0.5$ and $0.7$ fm, are chosen as typical parameters for constructing the $\alpha$--nucleus WS potentials. For each value of $a$, the potential depth $V_0$ is adjusted to satisfy the BS quantization condition in Eq.~\eqref{eq13}. In contrast, within the QCPM framework, the values of $a$ and $V_0$ are determined consistently using Eqs.~\eqref{eq8}, \eqref{eq12}, and \eqref{eq13}. These resulting potentials are then applied into Eqs.~\eqref{eq15}, \eqref{eq16}, and \eqref{eq17} to calculate $\alpha$ decay half-lives. The $\alpha$ preformation factors $S_c$ are taken from the generalized liquid drop model, evaluated using Eq.~(7) with the parameter set listed in Table~1 of Ref.~\cite{Deng21}. In the following, these preformation factors are denoted as GLDM. The analysis focuses on the half-lives of even--even Po isotopes with neutron numbers spanning the $N_p=126$ shell closure.

The logarithmic ratios are shown in Fig.~\ref{fig4_dt}. The blue and black curves represent the cases with the fixed values of $a=0.5$ and $0.7$ fm, respectively. The red curve denotes the results using our calculated diffuseness values, which fluctuate around 0.6 fm. These choices lead to the noticeably different logarithmic ratios. The calculated diffuseness values, which are close to the best-fit value \cite{Sun16,Buck95}, provide a reasonable description of the $\alpha$ decay half-lives. The corresponding logarithmic ratios lie between those of $a=0.5$ and 0.7 fm. Notably, the $\Delta {\rm log}_{10}(T_{1/2})$ curves exhibit significant fluctuations across $N_p=126$, possibly due to shell structure effects. These findings indicate that the calculated $\alpha$ decay half-lives are sensitive to the choice of diffuseness, particularly in the vicinity of the $N_p=126$ shell closure.

\subsection{$\alpha$ decay half-life}
\label{sec3b}

In this section, we calculate the $\alpha$ decay half-lives of heavy even--even nuclei with $140 \leq A_p \leq 260$, considering both the inclusion and exclusion of daughter-nucleus deformation. The WS potential with the depth $V_0$ and diffuseness $a$ from the QCPM is employed. The input data are the same as those used in Sec.~\ref{sec3a}. Throughout the calculations, the deformation parameters $\beta_2$ and $\beta_4$ are taken from Ref.~\cite{Moller16}.

In calculations that include deformation effects, the potential depth $V_0$ and diffuseness $a$ are first determined using Eqs.~\eqref{eq8}, \eqref{eq12}, and \eqref{eq13}. At each angle $\theta$, these two values are kept fixed, while a normalization factor $\lambda(\theta)$ is applied in the potential of Eq.~\eqref{eq3} to satisfy the BS condition. This procedure is mostly used in calculations that incorporate deformation effects \cite{Coba12}.  

\begin{figure}[htb]
	\centering
	\includegraphics[width=8.6cm]{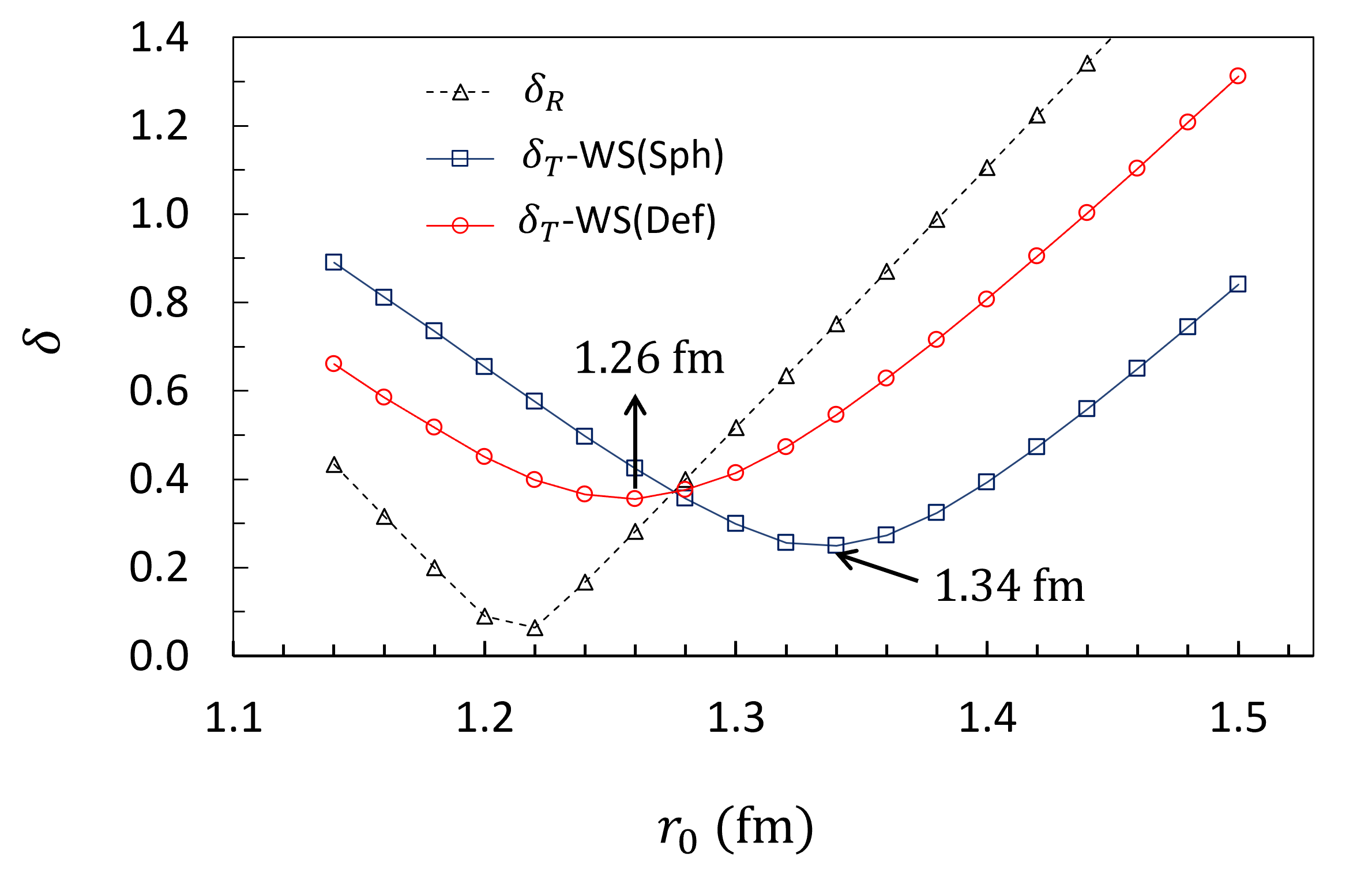}
	\caption{Root-mean-square (rms) deviations of the decimal logarithm of half-lives (\(\delta_T\)) and of radius (\(\delta_R\)). The solid curves with circles and squares denote \(\delta_T\) obtained from deformed and spherical calculations, respectively. The dashed curve with triangle markers corresponds to \(\delta_R\). The definitions of these deviations are given in Eqs.~\eqref{eq24} and \eqref{eq25}.}
	\label{fig5}
\end{figure}

We begin by examining the reliability of the radius parameter \(R_0\) in Eq.~\eqref{eq4} for the description of $\alpha$ decay half-lives. The reliability is evaluated by comparing the radius \(R_0\) with the conventional approximation \(R_\text{apx} = r_0 A_p^{1/3}\). In this analysis, the $\alpha$ decay half-lives are recalculated by replacing \(R_0\) with \(R_\text{apx}\), and the root-mean-square (rms) deviation of the decimal logarithm of the calculated half-lives relative to the experimental data, \(\delta_T \), is determined as a function of \(r_0\), as shown in Fig.~\ref{fig5}. The rms deviation is defined as
	\begin{equation}
	 \delta_T=\left[ \dfrac{1}{N_0}\sum^{N_0}_{i=1} \left( \log_{10}[T^\mathrm{cal}_{1/2}(i)/T^\mathrm{exp}_{1/2}(i)] \right)^2 \right]^{1/2}. \label{eq24}
	\end{equation}
Here, \(T^\mathrm{cal}_{1/2}\) and \(T^\mathrm{exp}_{1/2}\) denote the calculated and experimental half-lives, respectively. \(N_0=113\) is the total number of data points. The calculations are performed for \(r_0\) values ranging from 1.14 to 1.50~fm in steps of 0.02~fm. In Fig.~\ref{fig5}, the solid curve, marked with circles for deformed calculations and squares for those without deformation effects, represents the rms deviations of the logarithmic half-lives. The dashed curve with triangle markers shows the rms deviation $\delta_R$ between the approximate radius $R_\mathrm{apx}$ and the value of \(R_0\) in Eq.~\eqref{eq4}, given by
	\begin{equation}	
	  \delta_R = \left[ \dfrac{1}{N_0}\sum^{N_0}_{i=1} \left( R_\mathrm{apx}(i)-R_0(i) \right)^2 \right]^{1/2}. \label{eq25}
	\end{equation}	   

As shown in Fig.~\ref{fig5}, the minimum rms deviation of the logarithmic half-lives occurs at \(r_0\simeq1.34~\mathrm{fm}\) and \(1.26~\mathrm{fm}\) for the spherical and deformed calculations, respectively. In other words, accounting for nuclear deformation leads to a smaller optimal value of \(r_0\). The radius parameter $R_0$ in Eq.~\eqref{eq4}, corresponding to \(r_0\simeq1.22~\mathrm{fm}\), is close to the optimal value of \(r_0\simeq1.26~\mathrm{fm}\). In addition, the rms deviations, \(\delta_T\)-\(\mathrm{WS(Def)}\), obtained from using \(R_0\) and the optimal \(R_\mathrm{apx}\) differ by less than 13\(\%\). In the literature, values of \(r_0=1.07 ~\mathrm{and}~ 1.20 ~\mathrm{fm}\) are commonly adopted in studies of \(\alpha\) decay \cite{Ni09,Coba12}. These results support the reliability of the radius parameter in Eq.~\eqref{eq4} for the present framework. 

Figure \ref{fig5} also shows that the calculations, both with and without deformation effects, are highly sensitive to the choice of the radius parameter. In particular, when deformation is included, using \(r_0 < 1.28~\mathrm{fm}\) yields better agreement between the theoretical and experimental results compared to the spherical cases. In contrast, using \(r_0 \gtrsim 1.28~\mathrm{fm}\) leads to a noticeable deviation from the measured half-life data. Therefore, it should be noted that although a larger \(r_0 \sim 1.28-1.36~\mathrm{fm}\) seem to reproduce the experimental half-lives well in spherical calculations, it is likely unphysical since it leads to inconsistencies when deformation is considered.

\begin{figure}[hbt]
	\centering
	\includegraphics[width=8.6cm]{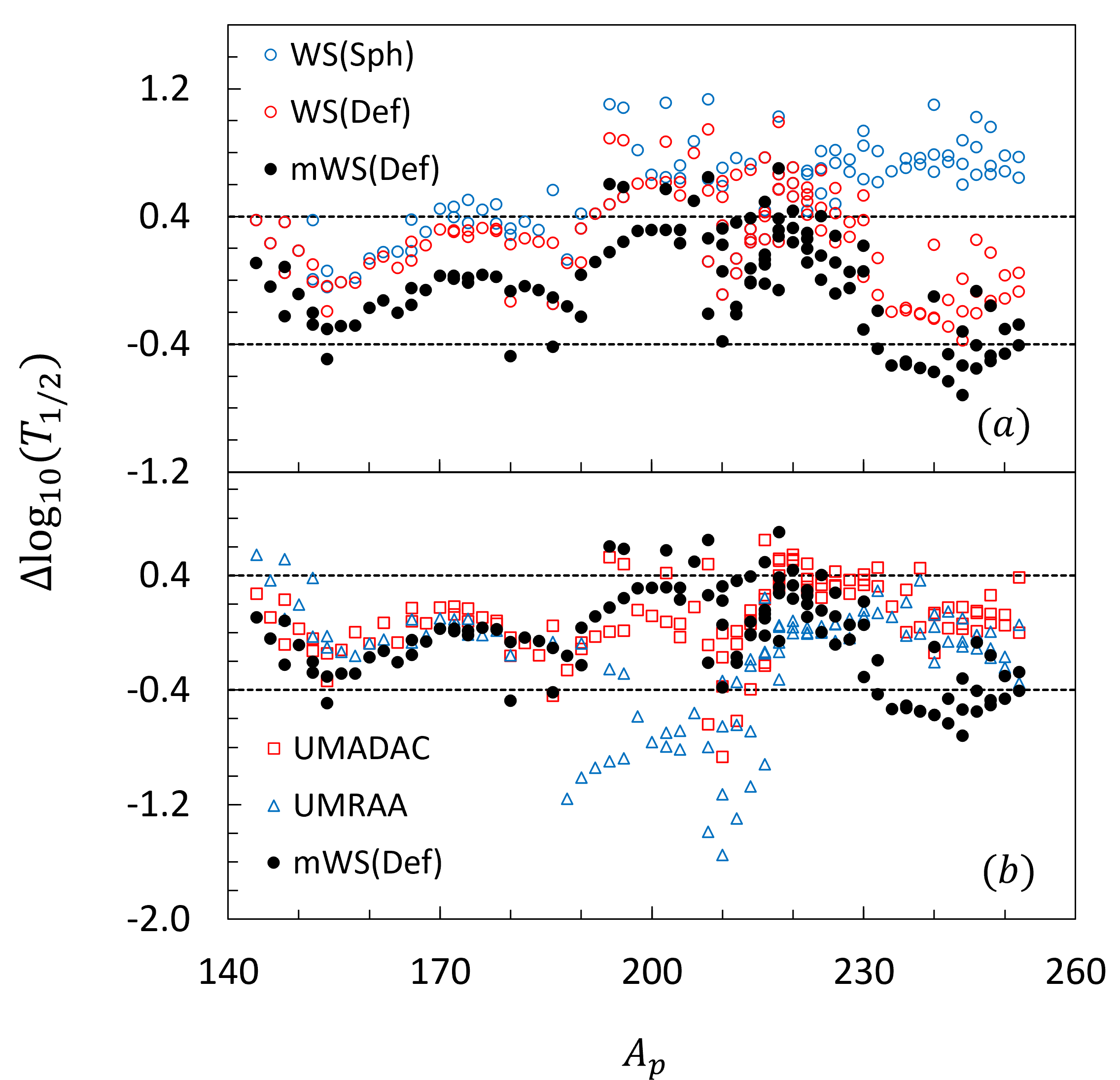}
	\caption{Comparison of logarithmic ratios of the theoretical $\alpha$ decay half-lives to the experimental data. (a) Results obtained using the WS potential in Eq.~\eqref{eq3} with (red open circles) and without (blue open circles) inclusion of daughter-nucleus deformation, and using the modified WS (mWS) potential including daughter-nucleus deformation (black solid points). (b)  Comparison of the mWS results with the analytical results of UMADAC \cite{Dens25} (red open squares) and UMRAA \cite{Dumi22} (blue open triangles). The dashed lines are drawn as guides to the eye.}
	\label{fig6}
\end{figure}

After using $R_\mathrm{apx}$ in the previous calculations, we now turn to $R_\mathrm{0}$ in Eq.~\eqref{eq4} to calculate the \(\alpha\) decay half-lives and compare the results with those obtained from two recent realistic analytical approaches, named UMADAC \cite{Dens25} and UMRAA \cite{Dumi22}. The theoretical and experimental $\alpha$ decay half-lives, along with the corresponding diffuseness values, are listed in Table~\ref{tb4}. For a clearer visualization of the calculated results, we present in Fig.~\ref{fig6} the logarithmic ratios of half-lives $\Delta {\rm log}_{10}(T_{1/2})$ as functions of mass number. In panel (a), it is observed that the inclusion of nuclear deformation (red open circles) significantly reduces $\Delta {\rm log}_{10}(T_{1/2})$ compared with the calculations without deformation (blue open circles). This improvement is particularly pronounced for heavy nuclei in the mass region $230 \leq A_p \leq 260$, where the theoretical predictions agree more closely with the experimental data. Furthermore, as shown in the second row of Table \ref{tb2}, incorporating deformation effects reduces the rms deviation from 0.608 to 0.407, comparable to the value of 0.461 from UMRAA in Ref.~\cite{Dumi22}. These results highlight the crucial role of nuclear deformation in studies of the $\alpha$ decay process, in agreement with the conclusions of Ref.~\cite{Coba12}. 

Although the WS potential, with the constrained parameters $V_0$ and $a$, provides a reasonable description of the $\alpha$ decay half-life data, its rms deviation remains relatively large, as shown in the second row of Table \ref{tb2}. Therefore, further improvement in the theoretical description is desirable. In general, one could independently adjust the values of $V_0$ and $a$ to better fit the data. However, such adjustments may lack physical significance, as these two parameters should be constrained by the relations given in Eqs.~\eqref{eq8} and \eqref{eq13}. 

In a microscopic description of $\alpha$ cluster formation inside the $^{212}$Po nucleus \cite{Xu16,Ropke14}, it has been shown that the $\alpha$ cluster only emerges at distances beyond a critical radius, where the baryon density is approximately 0.03 fm$^{-3}$. At smaller separations, the $\alpha$ cluster rapidly dissolves, with its four nucleons becoming nearly uncorrelated. This phenomenon, known as the nuclear medium effect, originates from Pauli blocking. Incorporating this nuclear medium effect into the double-folding model calculations leads to a strong reduction of the nuclear potential in the surface region (as shown in Fig.~3 in Ref.~\cite{Deng17}). This reduction has been shown to improve agreement between theoretical predictions and experimental $\alpha$ decay half-life data. 

Motivated by these studies \cite{Xu16,Ropke14,Deng17}, we propose a modification to the WS potential that approximates this surface suppression. This adjustment is expected to enhance the accuracy of the calculated half-lives. After careful testing, the modified WS potential is introduced as described below
 \begin{equation}
 	 	V^\text{m}_{N}(r) = -\frac{V_0}{2} \left[
 	\frac{1}{1 + \exp\!\left(x \right)} + \frac{\exp\!\left( x \right)}{\left[ 1 + \exp\!\left( x \right) \right]^2}
 	\right]
 	. \label{eq26}
 \end{equation}
Here, $x=(r - R_0)/a$. The factor of $1/2$ is included as a convention to ensure that the modified form retains the asymptotic behavior of the WS potential at large distances, $-V_0 \exp( -x )$. The potential depth $V_0$ in Eq.~\eqref{eq26}, which is obtained by solving the equation \( \dfrac{d V_{\rm eff}}{dr} \Big|_{r=R_B} = 0 \), is given by
\begin{equation}
	V_0 = \dfrac{Z a}{R_B^2} 
	\dfrac{\Big [ 1+\exp\!{\left [(R_B - R_0)/a \right ]} \Big ]^3}
	{\exp\!{\left [2(R_B - R_0)/a \right ] }}.\label{eq27}
\end{equation}

\begin{figure}[htb]
 	\centering
 	\includegraphics[width=8.6cm]{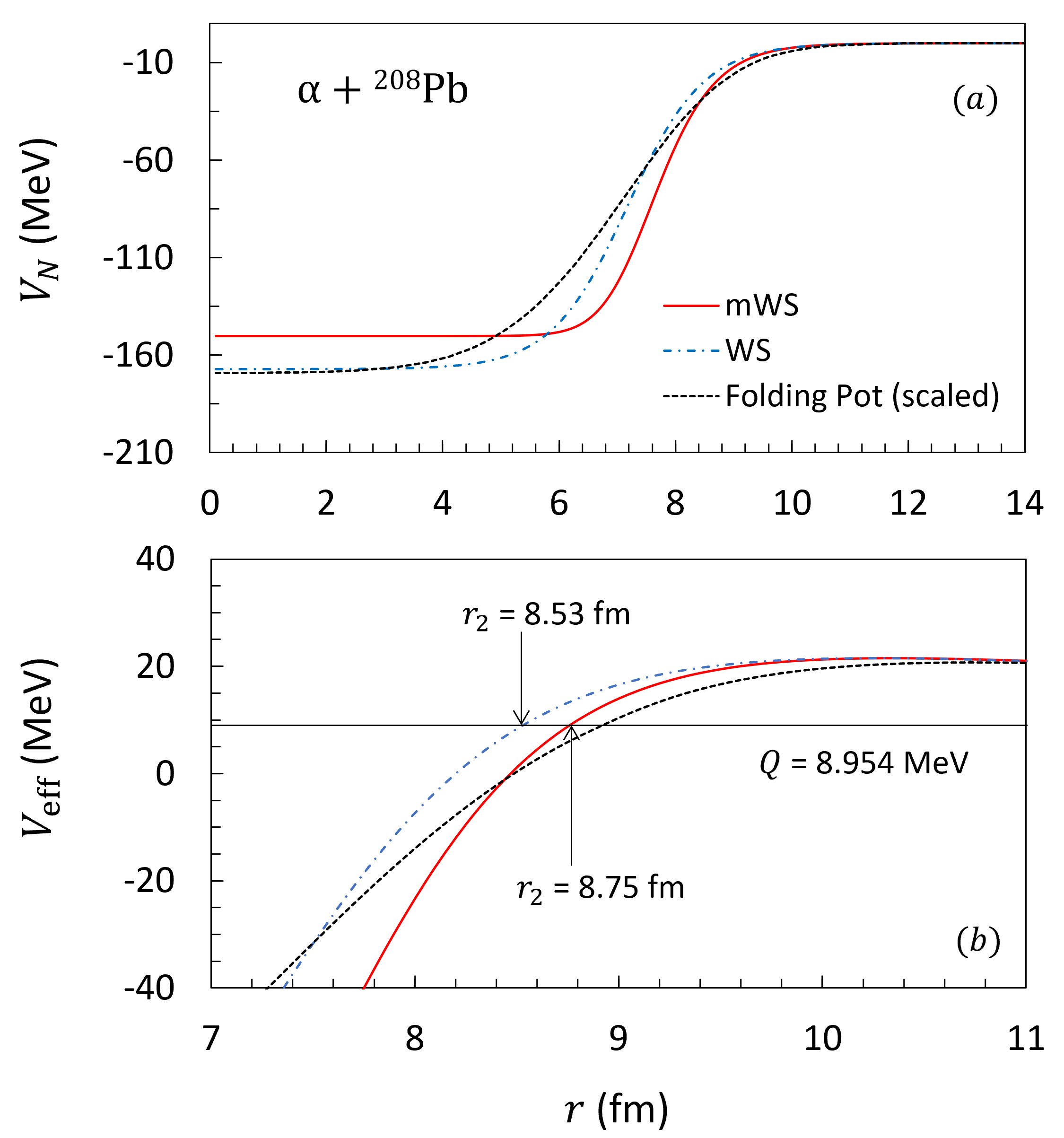}
 	\caption{(a) The spherical $\alpha+^{208}$Pb nuclear potential is calculated using the WS, mWS, and folding models. WS parameters in Eq.~\eqref{eq3} are obtained via the QCPM using Eqs.~\eqref{eq8}, \eqref{eq12}, and \eqref{eq13}. For the mWS potential, the same procedure is applied, but Eq.~\eqref{eq8} is replaced by Eq.~\eqref{eq27}. Details of the folding calculation are provided in Ref.~\cite{Chien22}. (b) The spherical effective potential $V_\text{eff}(r)$ in the Coulomb barrier region. $r_2$ denotes the second turning point, which separates the internal and external regions.}
 	\label{fig7}
 \end{figure}

To calculate the modified WS potential $V^\text{m}_N (r)$, the QCPM is applied to determine the parameters $V_0$ and $a$ using Eqs.~\eqref{eq12}, \eqref{eq13}, and \eqref{eq27}. In this procedure, $R_B$ in Eq.~\eqref{eq27} is determined by Eq.~\eqref{eq12} and the standard WS potential in Eq.~\eqref{eq13} is replaced by the modified form in Eq.~\eqref{eq26}. The resulting potential for the $\alpha+^{208}$Pb system (red solid curve) is presented in Fig.~\ref{fig7}, in comparison with the standard WS potential (blue dash-dotted curve) and the folding potential (black dotted curve, scaled to satisfy the BS condition as mentioned in Fig.~\ref{fig2_p}). Hereafter, the modified WS potential is referred to as mWS.
 
 As illustrated in panel (a) of Fig.~\ref{fig7}, the mWS potential closely follows the folding potential in the surface region, which plays a crucial role in accurately predicting $\alpha$ decay half-lives. This modification causes the nuclear potential to fall off more rapidly, shifting the second turning point $r_2$ from 8.53 to 8.75 fm. As a result, the Coulomb barrier becomes narrower, as shown in panel (b). 

Returning to panel (a) of Fig.~\ref{fig6}, we compare the logarithmic ratios of the theoretical results to the experimental half-life data. The use of the mWS potential incorporating nuclear deformation (black solid points) significantly improves the agreement between the theory and experiment. Although a small discrepancy remains in the mass region $230 \leq A_p \leq 260$, the results obtained with the mWS potential are generally consistent with the previous predictions \cite{Dens25,Dumi22}, as shown in panel (b).

The corresponding rms deviations are also compared with those obtained from the analytical models \cite{Dens25,Dumi22}, as presented in the second row of Table~\ref{tb2}. In Ref.~\cite{Dens25}, a systematic fitting procedure within the UMADAC model yields excellent agreement between the experimental and theoretical results, with an rms deviation of $\delta_T=0.283$. The realistic analytical approach (UMRAA) proposed in Ref.~\cite{Dumi22} gives an rms deviation of $\delta_T=0.461$. The use of the mWS potential results in an rms deviation of $\delta_T=0.323$, falling between those of the previous models. This analysis suggests that the modified WS potential in Eq.~\eqref{eq26} can be used as a reliable alternative for studying $\alpha$ decay half-lives. It also indicates that the diffuseness derived from the QCPM using Eqs.~\eqref{eq12}, \eqref{eq13}, and \eqref{eq27} is sufficient to yield reliable $\alpha$ decay half-life predictions.

\begin{table}[h]
	\centering
	\caption{Root-mean-square (rms) deviations \(\delta_T\) of the decimal logarithm of half-lives predicted by different potential models. $N_0$ is the number of data points used in Eq.~\eqref{eq24}. The results obtained using the WS potential with and without daughter-nucleus deformation are labeled as WS(Sph) and WS(Def), respectively. The modified WS potential defined in Eq.~\eqref{eq26}, which includes daughter-nucleus deformation, is denoted as mWS(Def). The three models for the $\alpha$ preformation factor $S_c$ in Eq.~\eqref{eq15}, namely GLDM, PKM, and SEM, are described in detail in the main text.
		\label{tb2}}
	\setlength{\tabcolsep}{2.0pt}
	\begin{tabular}{cccccc}
		\toprule
		\rule{0pt}{3.5ex} 
		$S_c$ &$\text{WS}$ &
		$\text{WS}$ & 
		$\text{mWS}$ &
		UMADAC & 
		UMRAA  \\ 
		model &$\text{(Sph)}$ &
		$\text{(Def)}$ & 
		$\text{(Def)}$ & \cite{Dens25}
		& \cite{Dumi22}  \\ \cline{2-5}
		&
		& 
		& $(N_0=113)$ &
		& 
		$(N_0=103)$ \rule{0pt}{3.5ex} \\
		\midrule
		GLDM \cite{Deng21}&0.608 & 0.407 & 0.323 & 0.283 & 0.461 \\
		\midrule
		~~PKM \cite{Sun25} &0.568 & 0.348 & 0.327 &   &   \\
		\midrule
		~~SEM \cite{Zhang17} &1.309 & 1.056 & 0.776 &   &    \\
		\bottomrule
	\end{tabular}
\end{table}

To strengthen our conclusions regarding the mWS potential, we also calculate the $\alpha$ decay half-lives using two additional, independently obtained $\alpha$ preformation factors. The first is based on the pick-up mechanism, with values taken from Table~II of Ref.~\cite{Sun25}, and is denoted as PKM. The second is derived from a semi-empirical formula for $\alpha$ decay half-lives using Eqs.~(6)--(8) of Ref.~\cite{Zhang17} and the parameters listed in Table~I therein. This approach is hereafter referred to as SEM. All the calculations are performed in the same manner as previously described, but with the GLDM values replaced by the PKM and SEM values, respectively. A comparison of results obtained with the three different $\alpha$ preformation factors is presented in Table \ref{tb2}.

As shown in Table \ref{tb2}, the use of the mWS potential with nuclear deformation yields the smallest rms deviation. This again confirms that the mWS potential provides better agreement between the theoretical predictions and the experimental $\alpha$ decay half-lives compared with the conventional WS potential. In addition, the obtained results demonstrate that the $\alpha$ preformation factor is strongly dependent on the model used for its determination, leading to considerable differences in the corresponding rms deviations. Therefore, independently developed microscopic models for the $\alpha$ preformation factor are desirable.

In addition, we compare the results obtained using the WS potential and the cosh-parameterized form \cite{Buck89,Buck93}. When using the same parameter set ($V_0$, $a$, and $R_0$), both potential types yield nearly identical $\alpha$ decay half-lives. This similarity arises because, in most practical cases, the $\alpha$--heavy nucleus potential satisfies the condition $R_0 \gg a$, under which the WS and cosh-parameterized forms become effectively equivalent \textcolor{black}{\cite{Buck77}}.

\subsection{Cluster decay half-life}
\label{sec3c} 
 
In cluster decay studies, reliable values of the potential depth $V_0$ and diffuseness $a$ are difficult to determine due to limited experimental data. Moreover, some global nucleus--nucleus potentials are too shallow to support the cluster-core quasi-bound state. To address these issues, we apply the QCPM to consistently determine $V_0$ and $a$. Specifically, Eqs.~\eqref{eq8}, \eqref{eq12}, and \eqref{eq13} are used to calculate these two parameters for the WS potential defined in Eq.~\eqref{eq3}. When applying the mWS potential, the same procedure is used, but Eq.~\eqref{eq8} is replaced by Eq.~\eqref{eq27}. 

In our calculations, the Coulomb barrier height \(V_B\) in Eq.~\eqref{eq12} is obtained from a global potential (GP) \cite{Broglia04}. Specifically, the GP is expressed in the WS form, with parameters taken from Eqs.~(30) and (33) on page 112 and Eqs.~(40), (41), (44), and (45) on page 114 therein. The nuclei considered in this analysis are listed in Table \ref{tb5}. The corresponding decay energies and half-life data are adopted from Refs.~\cite{Wang21,Kondev21}, respectively. The deformation parameters $\beta_2$ and $\beta_4$ of the daughter nuclei are taken from Ref.~\cite{Moller16}. The cluster preformation factors are taken from the eighth column of Table~I in Ref.~\cite{Qi23}. For the \(^{225}_{89}\!\mathrm{Ac}\) nucleus, the \(^{14}_{6}\mathrm{C}\) cluster preformation factor, not listed in Ref.~\cite{Qi23}, is adopted from Table~1 in Ref.~\cite{Dong09}.
 \begin{figure}[htb]
	\centering
	\includegraphics[width=8.6cm]{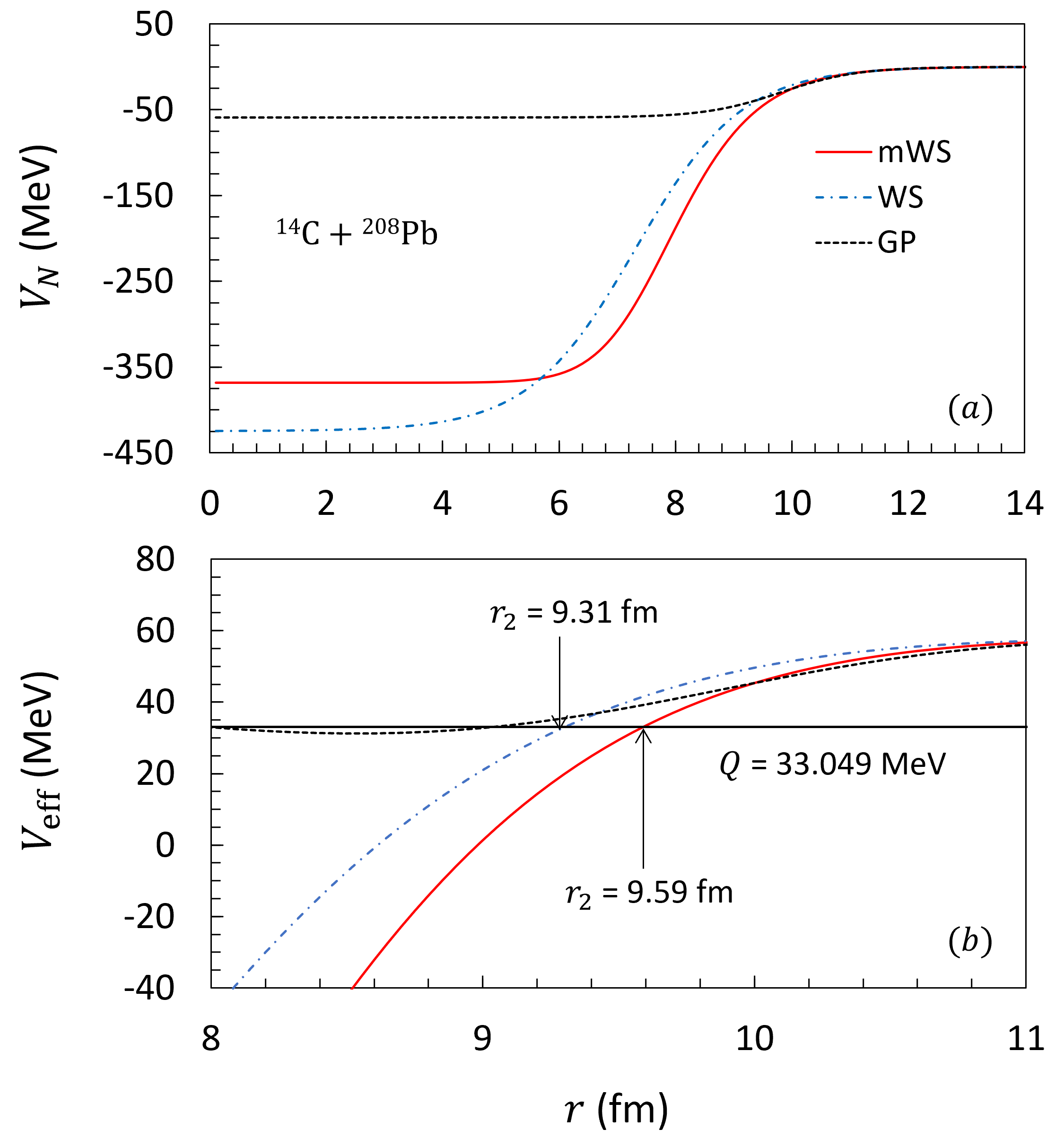}
	\caption{Same as Fig.~\ref{fig7}, but for the $^{14}\mathrm{C}+^{208}\!\mathrm{Pb}$ system, with the folding potential replaced by the GP potential, taken from Ref.~\cite{Broglia04}.}
	\label{fig8}
\end{figure}

Figure~\ref{fig8} presents a comparison of the WS, mWS, and GP potentials. In the surface region, where the penetration probability in Eq.~\eqref{eq17} is most sensitive, the mWS potential (red solid curve) follows the GP potential (black dotted curve) more closely than the WS potential (blue dash-dotted curve). At short distances, the mWS potential becomes deeper than the GP potential, appropriately accounting for the $^{14}\mathrm{C}+^{208}\!\mathrm{Pb}$ quasi-bound state.

\begin{figure}[htb]
	\centering
	\includegraphics[width=8.6cm]{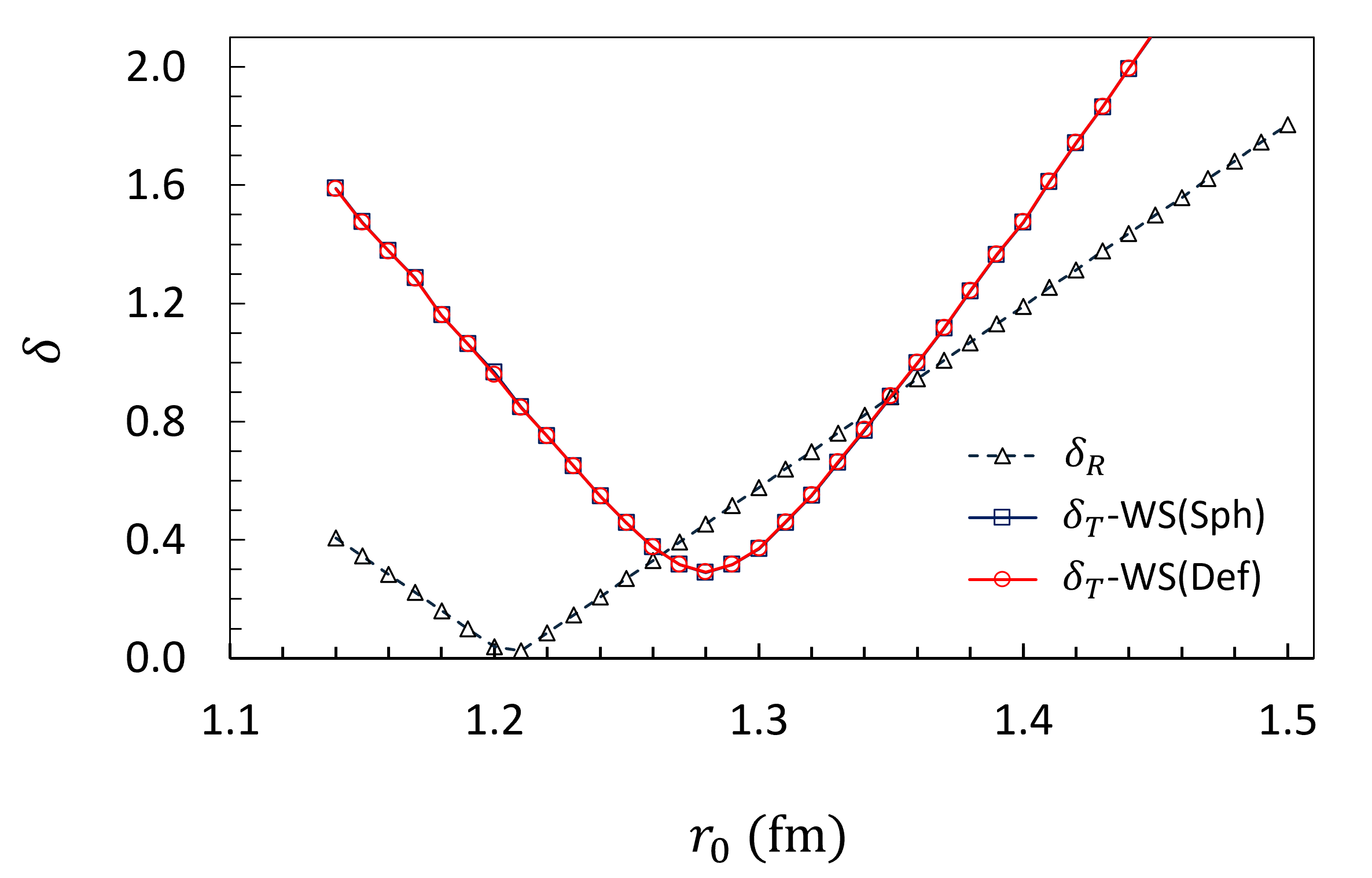}
	\caption{Same as Fig.~\ref{fig5} but for cluster decay.}
	\label{fig9}
\end{figure}

We next examine the reliability of the radius parameter in Eq.~\eqref{eq4} in describing cluster decay. This analysis is performed in the same manner as for the \(\alpha\) calculations but using a redefined radius \( R_\mathrm{apx} = r_0 A_d^{1/3} + \Delta\!R_\mathrm{apx}\) with \( \Delta\!R_\mathrm{apx} = r_0 (A_p^{1/3} - A_d^{1/3}) \). The definition of \(R_\mathrm{apx}\) not only accounts for the finite size of emitted clusters through \( \Delta\!R_\mathrm{apx} \) but also allows inclusion of deformation effects of daughter nuclei. We note that, in the absence of deformation, \( R_\mathrm{apx}\) reduces to \(r_0 A_p^{1/3}\). The corresponding results are summarized in Fig.~\ref{fig9}. 

As shown in Fig.~\ref{fig9}, the rms deviations \(\delta_T\) for the spherical and deformed calculations reach a minimum at \(r_0 \simeq 1.28~ \mathrm{fm}\). \textcolor{black}{The two sets of results are nearly identical, owing to the very small intrinsic deformation of the daughter nuclei (typically $\beta_2\simeq0.01$ \cite{Moller16})}. In the present work, the use of Eq.~\eqref{eq4} yields radii close to \(R_\mathrm{apx}\) at \(r_0 \simeq 1.21~ \mathrm{fm}\), where \(\delta_R\) attains its minimum. Although this \(r_0\) is slightly smaller than the optimum (\(\simeq 1.28~ \mathrm{fm}\)), the resulting potential remains reasonable for describing cluster decay, particularly in the surface region, as illustrated in Fig.~\ref{fig8}.

In this analysis, we also examine the reliability of the diffuseness values obtained from the QCPM. As shown in the sixth column of Table~\ref{tb5}, they span from \(0.91 ~\mathrm{to}~ 1.06~\mathrm{fm}\), exceeding the typical range \( 0.63-0.67 ~\mathrm{fm}\) extracted from elastic scattering analyses of heavy-ion systems \cite{Christ76,Bottoni12}. Although this difference may seem questionable, using the empirical diffuseness \(a\simeq0.67~\mathrm{fm}\) within the QCPM framework leads to a significantly larger rms deviation of the logarithmic half-lives. In particular, constraining \(a\) toward the empirical range requires increasing the radius parameter up to \(r_0\simeq 1.48~\mathrm{fm}\), which is considerably larger than the optimal value of \(\simeq1.28~\mathrm{fm}\). Such adjustment results in poor agreement with the cluster decay data, as illustrated in Fig.~\ref{fig9}. Even at the optimal radius, the corresponding diffuseness remains relatively large, ranging from \( 0.84~\mathrm{to}~0.98~\mathrm{fm} \).  

In addition, such enhanced diffuseness agrees with the observations from a systematic single-channel study of heavy-ion reactions, which found that \(a\sim 0.8-1.1 ~\mathrm{fm}\) is required to reproduce scattering data involving deformed targets \cite{Washiyama06}. This increase has been attributed to coupling effects \cite{Gasques07}. Taken together, these results suggest that the large diffuseness obtained from the QCPM is physically plausible for describing cluster decay.

We now return to the discussion of cluster decay half-lives calculated using the QCPM-based potentials, as listed in Table~\ref{tb5}. The corresponding rms deviations from the experimental data are presented in Table~\ref{tb3}. It should be noted that these calculations employ the radius parameter defined in Eq.~\eqref{eq4}, rather than \(R_\mathrm{apx}\) in the previous examination step.

As shown in Table~\ref{tb3}, the \(\delta_T\) values obtained from the spherical and deformed QCPM-based WS potentials are comparable to the UMRAA-based result \cite{Dumi22}, indicating that these potentials give a reasonable description of the cluster decay half-life data. The use of the mWS potential further improves the agreement. The rms deviation obtained with the mWS potential is smaller than that from UMRAA. These results confirm that the surface diffuseness and the potential depth consistently derived from Eqs.~\eqref{eq12}, \eqref{eq13}, and \eqref{eq27} are sufficiently reliable for describing the nuclear potential in both $\alpha$ and cluster decay studies.
\begin{table}[h]
	\centering
	\caption{Same as Table~\ref{tb2}, but for cluster decay half-lives.
		\label{tb3}}
	\setlength{\tabcolsep}{15pt}
		\begin{tabular}{cccc}
			\toprule
			\rule{0pt}{3.5ex} 
			$\text{WS}$ &
			$\text{WS}$ & 
			$\text{mWS}$ &
			UMRAA  \\ 
			$\text{(Sph)}$ &
			$\text{(Def)}$ & 
			$\text{(Def)}$ & \cite{Dumi22}  \\ 
			\cline{1-3}  
			& 
			$(N_0=17)$ &  & 	$(N_0=14)$\rule{0pt}{3.5ex} \\
			\midrule
			0.894 & 0.893 & 0.439 & 0.803 \\
			\bottomrule
		\end{tabular}
\end{table}

Using the mWS shape results in a more rapid fall-off of the nuclear potential in the surface region compared with the conventional WS form. This behavior may draw readers’ attention to the surface suppression of folding potentials by considering the nuclear medium effect \cite{Deng17,Deng19}, which arises from Pauli blocking in microscopic calculations \cite{Xu16,Ropke14}. Folding model studies incorporating this effect have been shown to significantly improve the description of experimental $\alpha$ decay data. Extending such investigations to cluster decay could therefore be of interest.

Based on Eqs.~\eqref{eq12}, \eqref{eq13}, and \eqref{eq27}, the QCPM provides a consistent framework for constructing the mWS nuclear potential without fitting the parameters $V_0$ and $a$. This method improves the predictive capability relative to empirically fitted parameterizations, particularly for cluster decay where experimental half-life data are often limited. Consequently, the mWS potential may serve as a complementary alternative to the semi-microscopic folding potential commonly used in such studies. 

\section{Conclusion}
\label{sec4}

In this work, we have proposed the QCPM approach to determine the potential depth $V_0$ and diffuseness $a$ of the WS potential, which are essential for studies of $\alpha$ and cluster decay. This approach provides a predictive framework in which the $V_0$ and $a$ parameters are calculated rather than obtained through fitting procedures. The reliability of the derived parameters is supported by the optical model analysis of elastic $\alpha+^{140}$Ce, $\alpha+^{144}$Sm, and $\alpha+^{208}$Pb scattering data.
The resulting WS potential is in good agreement with both the best-fit and double-folding models, particularly in the surface region, which plays a dominant role in determining the accuracy of $\alpha$ and cluster half-life calculations. 

The WS potential with the QCPM-derived $V_0$ and $a$ parameters has been applied within a semi-classical cluster framework to calculate the $\alpha$ and cluster decay half-lives. The model yields the results in reasonable agreement with the experimental data. The inclusion of daughter-nucleus deformation further improves the theoretical description. Our analysis also indicates that the choice of diffuseness $a$ has a significant impact on the calculated $\alpha$ decay half-lives. In addition, we have proposed a modified WS potential that effectively approximates surface suppression, as observed in folding calculations incorporating the nuclear medium effect. This modification improves the consistency between the theoretical predictions and the experimental $\alpha$ and cluster decay half-life data. 

\section*{Acknowledgments}
We are grateful to Professor Peter Mohr for valuable discussions on the BS condition. This research is funded by Vietnam National University, Ho Chi Minh City (VNU-HCM) under Grant Numbers B2025-18-01. 

\newpage
\appendix
\section{$\alpha$ and cluster decay half-lives}
\label{appa}
\begin{longtable*}{@{\extracolsep{\fill}} l l l l l l l l l l @{}} 
	\caption{$\alpha$ decay half-lives calculated using the present potentials are compared with experimental data \cite{Kondev21} and results obtained using two realistic analytical approaches from the literature \cite{Dens25,Dumi22}. $T_{1/2}^\text{WS} \text{(Sph)}$ and $T_{1/2}^\text{WS} \text{(Def)}$ represent the theoretical half-lives calculated using the Woods--Saxon (WS) potential without and with nuclear deformation effects, respectively. $T_{1/2}^\text{mWS} \text{(Def)}$ denotes the results obtained using the modified WS potential with deformation effects included. $a^\text{WS}$ and $a^\text{mWS}$ correspond to the diffuseness values of the WS and modified WS potentials, respectively. \textcolor{black}{The symbol “--” indicates nuclei not listed in Ref.~\cite{Dumi22}.}} 
	\label{tb4} \\ \toprule
	$\alpha$ & $Q$ \cite{Wang21} & $T_{1/2}^{\rm exp}$ \cite{Kondev21} & $a^\text{WS}$ & $a^\text{mWS}$ & $T_{1/2}^\text{WS} \text{(Sph)}$
	& $T_{1/2}^\text{WS} \text{(Def)}$ & $T_{1/2}^\text{mWS} \text{(Def)}$
	& $T_{1/2}$ \cite{Dumi22}  
	& $T_{1/2}$ \cite{Dens25} \\
	emitter & (MeV) & (s) & (fm) & (fm) &  (s)        
	& (s) & (s) & (s) & (s)  \\  \midrule 
	\endfirsthead
	\multicolumn{9}{c}
	{\tablename\ \thetable\ -- \textit{Continued}} \\ [5pt]
	\hline 
	$\alpha$ & $Q$ \cite{Wang21} & $T_{1/2}^\text{exp}$ \cite{Kondev21} &  $a^\text{WS}$ &  $a^\text{mWS}$ &  $T_{1/2}^\text{WS} \text{(Sph)} $  
	& $T_{1/2}^\text{WS} \text{(Def)}$ & $T_{1/2}^\text{mWS} \text{(Def)}$ 
	& $T_{1/2}$ \cite{Dumi22} 
	& $T_{1/2}$ \cite{Dens25} \\
	emitter & (MeV) & (s) & (fm) & (fm) & (s)         
	& (s) & (s) & (s) & (s)  \\  \midrule 
	\endhead
	\multicolumn{9}{r}{\textit{Continued on next page}} \\
	\endfoot
	\endlastfoot
	$^{144}_{60}$Nd& 1.901 & $7.227\times10^{22}$ &0.619& 0.553&
$1.725\times10^{23}$&	 $1.725\times10^{23}$   & $9.270\times10^{22}$
	&$2.523\times10^{23}$  	& $1.349\times10^{23}$ 
\rule{0pt}{10pt} \\
	$^{146}_{62}$Sm& 2.529 & $2.146\times10^{15}$ &0.618& 0.552&
$3.669\times10^{15}$&	$3.669\times10^{15}$	& $1.960\times10^{15}$
	&$4.989\times10^{15}$  & $2.754\times10^{15}$ 
\rule{0pt}{10pt} \\ 	
	$^{148}_{62}$Sm& 1.987 & $1.988\times10^{23}$ &0.614& 0.547&
$4.592\times10^{23}$&	$4.592\times10^{23}$ & $2.411\times10^{23}$
	&$6.501\times10^{23}$  & $3.388\times10^{23}$ 
	\rule{0pt}{10pt} \\
	$^{148}_{64}$Gd& 3.271 & $2.250\times10^{9}$ &0.618& 0.551&
$2.500\times10^{9}$&	$2.500\times10^{9}$	& $1.345\times10^{9}$ 
    &$2.818\times10^{9}$ & $1.862\times10^{9}$  
\rule{0pt}{10pt} \\
	$^{150}_{64}$Gd& 2.807 & $5.649\times10^{13}$ &0.616& 0.549&
$8.692\times10^{13}$&	$8.692\times10^{13}$	& $4.621\times10^{13}$
	&$8.892\times10^{13}$ & $6.026\times10^{13}$ 
\rule{0pt}{10pt} \\
	$^{152}_{64}$Gd& 2.204 & $3.408\times10^{21}$ &0.612& 0.546&
$8.133\times10^{21}$&	$4.279\times10^{21}$ & $2.137\times10^{21}$
	&$8.204\times10^{21}$ & $3.090\times10^{21}$ 
\rule{0pt}{10pt} \\
	$^{152}_{68}$Er& 4.934 & $1.144\times10^{1}$ &0.623& 0.555&
$1.164\times10^{1}$&	$1.124\times10^{1}$	& $6.023\times10^{0}$
	&$1.072\times10^{1}$ & $8.511\times10^{0}$ 
\rule{0pt}{10pt} \\ 
	$^{154}_{66}$Dy& 2.945 & $9.467\times10^{13}$ &0.614& 0.547&
$1.088\times10^{14}$&	$6.037\times10^{13}$	& $3.046\times10^{13}$
	&$8.974\times10^{13}$  & $4.365\times10^{13}$ 
\rule{0pt}{10pt} \\ 
	$^{154}_{70}$Yb& 5.474 & $4.417\times10^{-1}$ &0.622& 0.553&
$4.000\times10^{-1}$ &	$4.057\times10^{-1}$ & $2.185\times10^{-1}$ 
	&$3.483\times10^{-1}$ & $3.162\times10^{-1}$ 
\rule{0pt}{10pt} \\
	$^{156}_{72}$Hf& 6.026 & $2.300\times10^{-2}$ &0.620& 0.551&
$2.242\times10^{-2}$ &	$2.242\times10^{-2}$ & $1.190\times10^{-2}$ 
	&$1.683\times10^{-2}$  & $1.738\times10^{-2}$   
\rule{0pt}{10pt} \\ 
	$^{158}_{74}$W & 6.612 & $1.430\times10^{-3}$ &0.620& 0.550&
$1.483\times10^{-3}$ &	$1.380\times10^{-3}$ & $7.428\times10^{-4}$ 
	&$9.863\times10^{-4}$ 	& $1.445\times10^{-3}$ 
\rule{0pt}{10pt} \\ 
	$^{160}_{74}$W & 6.066 & $1.034\times10^{-1}$ &0.616& 0.546&
$1.420\times10^{-1}$ &	$1.317\times10^{-1}$ & $6.947\times10^{-2}$ 
	&$8.690\times10^{-2}$  & $8.710\times10^{-2}$ 
\rule{0pt}{10pt} \\
	$^{162}_{76}$Os& 6.768 & $2.100\times10^{-3}$ &0.617& 0.547&
$3.148\times10^{-3}$ &	$2.951\times10^{-3}$ & $1.566\times10^{-3}$ 
	&$1.884\times10^{-3}$ & $2.455\times10^{-3}$ 
\rule{0pt}{10pt} \\
	$^{164}_{76}$Os& 6.479 & $2.188\times10^{-2}$ &0.616& 0.547&
$3.304\times10^{-2}$ &	$2.614\times10^{-2}$ & $1.366\times10^{-2}$ 
	&  -- & $1.862\times10^{-2}$ 
\rule{0pt}{10pt} \\
	$^{166}_{76}$Os& 6.143 & $2.566\times10^{-1}$ &0.615& 0.546&
$6.149\times10^{-1}$ &	$4.466\times10^{-1}$	& $2.287\times10^{-1}$ 
	& $3.177\times10^{-1}$ & $3.090\times10^{-1}$ 
\rule{0pt}{10pt} \\
	$^{166}_{78}$Pt& 7.292 & $2.940\times10^{-4}$ &0.618& 0.548&
$4.495\times10^{-4}$ &	$3.923\times10^{-4}$ & $2.064\times10^{-4}$
	&$2.512\times10^{-4}$ & $4.365\times10^{-4}$ 
\rule{0pt}{10pt} \\
	$^{168}_{78}$Pt& 6.990 & $2.020\times10^{-3}$ &0.617& 0.547&
$4.071\times10^{-3}$ &	$3.361\times10^{-3}$	& $1.752\times10^{-3}$ 
	&$1.923\times10^{-3}$ & $2.344\times10^{-3}$ 
\rule{0pt}{10pt} \\
	$^{170}_{78}$Pt& 6.707 & $1.393\times10^{-2}$ &0.617& 0.547&
$3.908\times10^{-2}$ &	$2.908\times10^{-2}$	& $1.490\times10^{-2}$ 
	&$1.754\times10^{-2}$  & $2.089\times10^{-2}$ 
\rule{0pt}{10pt} \\
	$^{172}_{78}$Pt& 6.463 & $1.017\times10^{-1}$ &0.617& 0.547&
$2.929\times10^{-1}$ &	$2.043\times10^{-1}$	& $1.039\times10^{-1}$
	&$1.265\times10^{-1}$  & $1.549\times10^{-1}$
\rule{0pt}{10pt} \\
	$^{172}_{80}$Hg& 7.524 & $2.310\times10^{-4}$ &0.619& 0.548&
$5.743\times10^{-4}$ &	$4.737\times10^{-4}$	& $2.467\times10^{-4}$ 
	&$2.582\times10^{-4}$ & $3.090\times10^{-4}$ 
\rule{0pt}{10pt} \\
	$^{174}_{78}$Pt& 6.183 & $1.151\times10^{0}$ &0.618& 0.547&
$3.663\times10^{0}$&	$2.369\times10^{0}$ & $1.192\times10^{0}$ 
	&$1.432\times10^{0}$  & $1.698\times10^{0}$ 
\rule{0pt}{10pt} \\
	$^{174}_{80}$Hg& 7.233 & $2.000\times10^{-3}$ &0.619& 0.547&
$4.565\times10^{-3}$ &	$3.746\times10^{-3}$ & $1.928\times10^{-3}$ 
	&$1.945\times10^{-3}$  & $2.399\times10^{-3}$ 
\rule{0pt}{10pt} \\
	$^{176}_{80}$Hg& 6.897 & $2.256\times10^{-2}$ &0.618& 0.547&
$6.246\times10^{-2}$ &	$4.817\times10^{-2}$	& $2.444\times10^{-2}$ 
	&$2.178\times10^{-2}$ & $2.884\times10^{-2}$ 
\rule{0pt}{10pt} \\
	$^{178}_{80}$Hg& 6.577 & $2.994\times10^{-1}$ &0.617& 0.546&
$8.962\times10^{-1}$ &	$6.271\times10^{-1}$ & $3.142\times10^{-1}$  
	&$3.090\times10^{-1}$  & $3.631\times10^{-1}$  
\rule{0pt}{10pt} \\
	$^{178}_{82}$Pb& 7.789 & $2.500\times10^{-4}$ &0.621& 0.549&
$5.687\times10^{-4}$ &	$5.099\times10^{-4}$ & $2.636\times10^{-4}$ 
	&  --	 & $2.884\times10^{-4}$ 
\rule{0pt}{10pt} \\
	$^{180}_{74}$W & 2.515 & $5.018\times10^{25}$ &0.606& 0.538&
$1.057\times10^{26}$&	$3.697\times10^{25}$ & $1.683\times10^{25}$
	&  --  & $3.467\times10^{25}$ 
\rule{0pt}{10pt} \\
	$^{180}_{82}$Pb& 7.419 & $4.100\times10^{-3}$ &0.619& 0.548&
$7.890\times10^{-3}$ &	$6.891\times10^{-3}$ & $3.518\times10^{-3}$ 
	&$2.851\times10^{-3}$ & $3.802\times10^{-3}$ 
\rule{0pt}{10pt} \\
	$^{182}_{82}$Pb& 7.066 & $5.500\times10^{-2}$ &0.618& 0.547&
$1.285\times10^{-1}$ &	$1.005\times10^{-1}$ & $5.065\times10^{-2}$ 
	& --  & $4.677\times10^{-2}$ 
\rule{0pt}{10pt} \\
	$^{184}_{82}$Pb& 6.774 & $6.125\times10^{-1}$ &0.618& 0.547&
$1.268\times10^{0}$ &	$1.069\times10^{0}$ & $5.340\times10^{-1}$ 
	& --  & $4.266\times10^{-1}$ 
\rule{0pt}{10pt} \\
	$^{186}_{76}$Os& 2.821 & $6.311\times10^{22}$ &0.609& 0.540&
$2.320\times10^{23}$ &	$1.083\times10^{23}$	& $4.934\times10^{22}$ 
	&$5.420\times10^{22}$ & $7.079\times10^{22}$ 
\rule{0pt}{10pt} \\
	$^{186}_{84}$Po& 8.501 & $3.400\times10^{-5}$ &0.614& 0.544&
$2.429\times10^{-5}$ &	$2.414\times10^{-5}$ & $1.308\times10^{-5}$
	& -- & $1.230\times10^{-5}$ 
\rule{0pt}{10pt} \\
	$^{188}_{84}$Po& 8.082 & $2.700\times10^{-4}$ &0.612& 0.543&
$3.642\times10^{-4}$ &	$3.458\times10^{-4}$ & $1.856\times10^{-4}$ 
	&$1.875\times10^{-5}$ & $1.479\times10^{-4}$ 
\rule{0pt}{10pt} \\
	$^{190}_{78}$Pt& 3.269 & $1.524\times10^{19}$ &0.608& 0.538&
$3.991\times10^{19}$&	$1.978\times10^{19}$	& $9.013\times10^{18}$ 
	&$1.285\times10^{19}$  & $1.175\times10^{19}$ 
\rule{0pt}{10pt} \\
	$^{190}_{84}$Po& 7.693 & $2.450\times10^{-3}$ &0.611& 0.542&
$5.176\times10^{-3}$ &	$5.176\times10^{-3}$ & $2.655\times10^{-3}$ 
	&$2.393\times10^{-4}$ & $2.089\times10^{-3}$ 
\rule{0pt}{10pt} \\
	$^{192}_{84}$Po& 7.320 & $3.220\times10^{-2}$ &0.610& 0.541&
$8.402\times10^{-2}$ &	$8.402\times10^{-2}$ & $4.189\times10^{-2}$ 
	&$3.673\times10^{-3}$  & $3.020\times10^{-2}$ 
\rule{0pt}{10pt} \\
	$^{194}_{84}$Po& 6.987 & $3.920\times10^{-1}$ &0.609& 0.540&
$1.174\times10^{0}$ &	$1.174\times10^{0}$ & $5.895\times10^{-1}$ 
	&$4.943\times10^{-2}$ & $3.981\times10^{-1}$ 
\rule{0pt}{10pt} \\
	$^{194}_{86}$Rn& 7.862 & $7.800\times10^{-4}$ &0.608& 0.539&
$9.914\times10^{-3}$ &	$6.072\times10^{-3}$	& $3.118\times10^{-3}$ 
	&$4.345\times10^{-4}$ & $2.630\times10^{-3}$ 
\rule{0pt}{10pt} \\
	$^{196}_{84}$Po& 6.658 & $5.589\times10^{0}$ &0.609& 0.540&
$1.992\times10^{1}$&	$1.992\times10^{1}$	& $1.042\times10^{1}$ 
	&$7.925\times10^{-1}$ & $6.166\times10^{0}$ 
\rule{0pt}{10pt} \\
	$^{196}_{86}$Rn& 7.617 & $4.700\times10^{-3}$ &0.609& 0.540&
$5.687\times10^{-2}$ &	$3.549\times10^{-2}$	& $1.802\times10^{-2}$ 
	&$2.443\times10^{-3}$  & $1.413\times10^{-2}$ 
\rule{0pt}{10pt} \\
	$^{198}_{86}$Rn& 7.349 & $6.925\times10^{-2}$ &0.608& 0.539&
$4.538\times10^{-1}$ &	$2.797\times10^{-1}$	& $1.416\times10^{-1}$ 
	&$1.799\times10^{-2}$  & $1.000\times10^{-1}$ 
\rule{0pt}{10pt} \\
	$^{200}_{86}$Rn& 7.043 & $1.185\times10^{0}$ &0.608& 0.539&
$5.433\times10^{0}$&	$4.820\times10^{0}$	& $2.445\times10^{0}$
	&$2.042\times10^{-1}$  & $1.549\times10^{0}$ 
\rule{0pt}{10pt} \\
	$^{202}_{86}$Rn& 6.774 & $1.244\times10^{1}$ &0.608& 0.539&
$5.514\times10^{1}$&	$5.115\times10^{1}$	 & $2.576\times10^{1}$
	&$2.000\times10^{0}$  & $1.479\times10^{1}$ 
\rule{0pt}{10pt} \\
	$^{202}_{88}$Ra& 7.880 & $4.100\times10^{-3}$ &0.609& 0.540&
$5.297\times10^{-2}$ &	$3.036\times10^{-2}$	& $1.535\times10^{-2}$
	&$8.204\times10^{-4}$  & $1.072\times10^{-2}$ 
\rule{0pt}{10pt} \\
	$^{204}_{86}$Rn& 6.547 & $1.029\times10^{2}$ &0.609& 0.540&
$4.501\times10^{2}$ &	$4.241\times10^{2}$	& $2.122\times10^{2}$ 
	&$1.570\times10^{1}$  & $9.550\times10^{1}$   
\rule{0pt}{10pt} \\
	$^{204}_{88}$Ra& 7.637 & $6.000\times10^{-2}$ &0.610& 0.540&
$3.145\times10^{-1}$ &	$2.041\times10^{-1}$ & $1.021\times10^{-1}$ 	
	&$1.242\times10^{-2}$ & $6.918\times10^{-2}$ 
\rule{0pt}{10pt} \\
	$^{206}_{88}$Ra& 7.415 & $2.400\times10^{-1}$ &0.611& 0.542&
$1.788\times10^{0}$&	$1.500\times10^{0}$	& $7.525\times10^{-1}$ 
	&$6.622\times10^{-2}$ & $3.631\times10^{-1}$  
\rule{0pt}{10pt} \\
	$^{208}_{84}$Po& 5.216 & $9.145\times10^{7}$  &0.614& 0.545&
$1.201\times10^{8}$&	$1.201\times10^{8}$	& $5.653\times10^{7}$ 
	&$3.733\times10^{6}$ 	& $2.089\times10^{7}$  
\rule{0pt}{10pt} \\ 
	$^{208}_{88}$Ra& 7.273 & $1.276\times10^{0}$ &0.614& 0.544&
$5.494\times10^{0}$&	$4.666\times10^{0}$	& $2.338\times10^{0}$
	&$2.028\times10^{-1}$  & $1.047\times10^{0}$ \rule{0pt}{10pt} \\
	$^{208}_{90}$Th& 8.202 & $2.400\times10^{-3}$ &0.612& 0.542&
$3.261\times10^{-2}$ &	$2.116\times10^{-2}$ & $1.061\times10^{-2}$ 
	& -- & $7.244\times10^{-3}$ 
\rule{0pt}{10pt} \\
	$^{210}_{84}$Po& 5.408 & $1.196\times10^{7}$ &0.620& 0.550&
$9.747\times10^{6}$&	$9.747\times10^{6}$ & $4.951\times10^{6}$ 
	&$3.357\times10^{5}$  	& $1.622\times10^{6}$ 
\rule{0pt}{10pt} \\ 
	$^{210}_{86}$Rn& 6.159 & $9.000\times10^{3}$ &0.616& 0.546&
$1.989\times10^{4}$&	$1.989\times10^{4}$	& $1.022\times10^{4}$
	&$6.714\times10^{2}$  & $3.802\times10^{3}$ 
\rule{0pt}{10pt} \\ 
	$^{210}_{88}$Ra& 7.151 & $4.000\times10^{0}$ &0.616& 0.545&
$1.562\times10^{1}$&	$1.335\times10^{1}$ & $6.681\times10^{0}$ 	
	&$1.845\times10^{0}$ & $2.692\times10^{0}$ 
\rule{0pt}{10pt} \\
	$^{210}_{90}$Th& 8.069 & $1.600\times10^{-2}$ &0.615& 0.544&
$8.088\times10^{-2}$ &	$6.711\times10^{-2}$ & $3.370\times10^{-2}$  
	&$3.540\times10^{-3}$ & $1.585\times10^{-2}$  
\rule{0pt}{10pt} \\
	$^{212}_{86}$Rn& 6.385 & $1.434\times10^{3}$ &0.623& 0.552&
$1.969\times10^{3}$&	$1.969\times10^{3}$	& $9.752\times10^{2}$ 
	&$7.211\times10^{1}$  & $3.467\times10^{2}$ 
\rule{0pt}{10pt} \\ 
	$^{212}_{90}$Th& 7.958 & $3.170\times10^{-2}$ &0.617& 0.546&
$1.848\times10^{-1}$ &	$1.457\times10^{-1}$	& $7.289\times10^{-2}$ 
	&$7.211\times10^{-3}$	& $3.236\times10^{-2}$ 
\rule{0pt}{10pt} \\
	$^{214}_{88}$Ra& 7.273 & $2.438\times10^{0}$ &0.625& 0.554&
$5.099\times10^{0}$&	$5.099\times10^{0}$ & $2.899\times10^{0}$  
	&$2.061\times10^{-1}$ & $9.772\times10^{-1}$  
\rule{0pt}{10pt} \\ 
	$^{214}_{90}$Th& 7.827 & $8.700\times10^{-2}$ &0.619& 0.548&
$4.650\times10^{-1}$ &	$4.282\times10^{-1}$ & $2.141\times10^{-1}$ 
	&$1.786\times10^{-2}$  & $1.000\times10^{-1}$  
\rule{0pt}{10pt} \\
	$^{216}_{90}$Th& 8.072 & $2.628\times10^{-2}$ &0.626& 0.554&
$7.238\times10^{-2}$ &	$7.042\times10^{-2}$ & $3.554\times10^{-2}$ 
	&$3.177\times10^{-3}$  & $1.549\times10^{-2}$ 
\rule{0pt}{10pt} \\
	$^{216}_{92}$U & 8.531 & $6.900\times10^{-3}$ &0.619& 0.547&
$1.900\times10^{-2}$ &	$1.738\times10^{-2}$ & $8.704\times10^{-3}$
	& -- & $4.266\times10^{-3}$
\rule{0pt}{10pt} \\
	$^{218}_{92}$U & 8.775 & $3.540\times10^{-4}$ &0.626& 0.554&
$3.755\times10^{-3}$ &	$3.482\times10^{-3}$ & $1.781\times10^{-3}$ 
	&$1.663\times10^{-4}$  & $8.913\times10^{-4}$ 
\rule{0pt}{10pt} \\ 
	$^{212}_{84}$Po& 8.954 & $2.944\times10^{-7}$ &0.654& 0.582&
$3.252\times10^{-7}$ &	$3.252\times10^{-7}$ & $1.810\times10^{-7}$ 
	&$1.334\times10^{-7}$  	& $2.455\times10^{-7}$ 
\rule{0pt}{10pt} \\ 
	$^{214}_{84}$Po& 7.834 & $1.635\times10^{-4}$ &0.640& 0.569&
$2.832\times10^{-4}$ &	$2.832\times10^{-4}$ & $1.567\times10^{-4}$
	&$1.067\times10^{-4}$ 	& $2.344\times10^{-4}$ 
\rule{0pt}{10pt} \\ 
	$^{214}_{86}$Rn& 9.208 & $2.590\times10^{-7}$ &0.647& 0.576&
$4.681\times10^{-7}$ &	$4.681\times10^{-7}$ & $2.536\times10^{-7}$ 
	&$1.528\times10^{-7}$ & $3.162\times10^{-7}$ 
\rule{0pt}{10pt} \\
	$^{216}_{84}$Po& 6.906 & $1.440\times10^{-1}$ &0.628& 0.559&
$2.608\times10^{-1}$ &	$2.608\times10^{-1}$ & $1.374\times10^{-1}$ 
	&$1.030\times10^{-1}$	& $2.630\times10^{-1}$ 
\rule{0pt}{10pt} \\ 
	$^{216}_{86}$Rn& 8.198 & $2.900\times10^{-5}$ &0.635& 0.565&
$1.711\times10^{-4}$ &	$1.711\times10^{-4}$	& $9.021\times10^{-5}$ 
	& $5.093\times10^{-5}$ & $1.288\times10^{-4}$ 
\rule{0pt}{10pt} \\ 
	$^{216}_{88}$Ra& 9.526 & $1.720\times10^{-7}$ &0.642& 0.571&
$4.592\times10^{-7}$ &	$4.592\times10^{-7}$ & $2.485\times10^{-7}$ 
	&$1.265\times10^{-7}$  & $2.951\times10^{-7}$ 
\rule{0pt}{10pt} \\
	$^{218}_{84}$Po& 6.115 & $1.859\times10^{2}$ &0.619& 0.551&
$3.241\times10^{2}$&	$3.241\times10^{2}$ & $1.620\times10^{2}$ 
	&$1.361\times10^{2}$	& $3.802\times10^{2}$ 
\rule{0pt}{10pt} \\ 
	$^{218}_{86}$Rn& 7.262 & $3.375\times10^{-2}$ &0.624& 0.555&
$1.252\times10^{-1}$ &	$1.252\times10^{-1}$ & $6.373\times10^{-2}$ 
	&$3.715\times10^{-2}$ & $1.072\times10^{-1}$ 
\rule{0pt}{10pt} \\ 
	$^{218}_{88}$Ra& 8.540 & $2.591\times10^{-5}$ &0.630& 0.560&
$1.199\times10^{-4}$ &	$1.199\times10^{-4}$ & $6.318\times10^{-5}$ 
	&$2.924\times10^{-5}$  & $8.511\times10^{-5}$ 
\rule{0pt}{10pt} \\ 
	$^{218}_{90}$Th& 9.849 & $1.220\times10^{-7}$ &0.636& 0.565&
$4.557\times10^{-7}$ &	$4.557\times10^{-7}$ & $2.529\times10^{-7}$  
	&$1.038\times10^{-7}$ & $2.754\times10^{-7}$  
\rule{0pt}{10pt} \\
	$^{220}_{86}$Rn& 6.405 & $5.560\times10^{1}$ &0.614& 0.546&
$1.860\times10^{2}$&	$1.860\times10^{2}$ & $9.620\times10^{1}$ 
	&$6.067\times10^{1}$	& $1.950\times10^{2}$ 
\rule{0pt}{10pt} \\
	$^{220}_{88}$Ra& 7.594 & $1.810\times10^{-2}$ &0.619& 0.550&
$7.331\times10^{-2}$ &	$7.331\times10^{-2}$	& $3.860\times10^{-2}$ 
	&$1.799\times10^{-2}$  & $5.754\times10^{-2}$ 
\rule{0pt}{10pt} \\ 
	$^{220}_{90}$Th& 8.973 & $1.020\times10^{-5}$ &0.627& 0.557&
$5.205\times10^{-5}$ &	$5.205\times10^{-5}$ & $2.787\times10^{-5}$ 
	&$1.233\times10^{-5}$ & $3.311\times10^{-5}$ 
\rule{0pt}{10pt} \\ 
	$^{222}_{86}$Rn& 5.590 & $3.302\times10^{5}$ &0.605& 0.538&
$9.028\times10^{5}$&	$8.529\times10^{5}$ & $4.270\times10^{5}$ 
	&$3.573\times10^{5}$ 	& $1.000\times10^{6}$ 
\rule{0pt}{10pt} \\ 
	$^{222}_{88}$Ra& 6.678 & $3.360\times10^{1}$ &0.609& 0.541&
$1.547\times10^{2}$&	$1.048\times10^{2}$ & $5.302\times10^{1}$ 
	&$3.350\times10^{1}$  & $7.943\times10^{1}$ 
\rule{0pt}{10pt} \\ 
	$^{222}_{90}$Th& 8.133 & $2.240\times10^{-3}$ &0.618& 0.549&
$1.084\times10^{-2}$ &	$8.555\times10^{-3}$ & $4.443\times10^{-3}$ 
	&$2.203\times10^{-3}$  	& $4.677\times10^{-3}$ 
\rule{0pt}{10pt} \\ 
	$^{222}_{92}$U & 9.481 & $4.700\times10^{-6}$ &0.625& 0.554&
$1.622\times10^{-5}$ &	$1.622\times10^{-5}$ & $8.546\times10^{-6}$
 	& -- & $9.333\times10^{-6}$
\rule{0pt}{10pt} \\ 
	$^{224}_{88}$Ra& 5.789 & $3.138\times10^{5}$ &0.599& 0.533&
$1.100\times10^{6}$&	$6.443\times10^{5}$ & $3.165\times10^{5}$ 
	&$3.133\times10^{5}$ & $6.761\times10^{5}$ 
\rule{0pt}{10pt} \\ 
	$^{224}_{90}$Th& 7.299 & $1.040\times10^{0}$ &0.609& 0.541&
$5.232\times10^{0}$&	$2.958\times10^{0}$ & $1.485\times10^{0}$ 
	&$1.054\times10^{0}$  & $1.820\times10^{0}$ 
\rule{0pt}{10pt} \\ 
	$^{224}_{92}$U & 8.628 & $3.960\times10^{-4}$ &0.616& 0.546&
$2.557\times10^{-3}$ &	$1.942\times10^{-3}$ & $1.003\times10^{-3}$ 
	&$4.395\times10^{-4}$  	& $9.550\times10^{-4}$ 
\rule{0pt}{10pt} \\
	$^{226}_{88}$Ra& 4.871 & $5.049\times10^{10}$ &0.589& 0.523&
$1.517\times10^{11}$&	$8.713\times10^{10}$ & $4.186\times10^{10}$ 
	&$5.781\times10^{10}$  & $1.349\times10^{11}$ 
\rule{0pt}{10pt} \\  
	$^{226}_{90}$Th& 6.453 & $1.842\times10^{3}$ &0.600& 0.533&
$1.003\times10^{4}$&	$4.856\times10^{3}$ & $2.385\times10^{3}$ 
	&$2.133\times10^{3}$  & $3.890\times10^{3}$ 
\rule{0pt}{10pt} \\ 
	$^{226}_{92}$U& 7.701 & $2.690\times10^{-1}$ &0.606& 0.538&
$1.761\times10^{0}$&	$1.021\times10^{0}$ & $5.105\times10^{-1}$ 
	&$2.355\times10^{-1}$ & $5.754\times10^{-1}$ 
\rule{0pt}{10pt} \\ 
	$^{228}_{90}$Th& 5.520 & $6.035\times10^{7}$ &0.590& 0.524&
$2.895\times10^{8}$&	$1.129\times10^{8}$ & $5.407\times10^{7}$ 
		&$7.534\times10^{7}$ & $1.413\times10^{8}$ 
\rule{0pt}{10pt} \\ 
	$^{228}_{92}$U& 6.800 & $5.600\times10^{2}$ &0.597& 0.530&
$3.197\times10^{3}$&	$1.300\times10^{3}$ & $6.350\times10^{2}$ 
	& $5.117\times10^{2}$ & $1.047\times10^{3}$ 
\rule{0pt}{10pt} \\
	$^{230}_{90}$Th& 4.770 & $2.379\times10^{12}$ &0.582& 0.517&
$1.027\times10^{13}$&	$2.502\times10^{12}$ & $1.168\times10^{12}$ 
	&$3.396\times10^{12}$  & $5.129\times10^{12}$ 
\rule{0pt}{10pt} \\  
	$^{230}_{92}$U& 5.992 & $1.748\times10^{6}$ &0.588& 0.522&
$1.223\times10^{7}$&	$4.158\times10^{6}$ & $1.984\times10^{6}$ 
	&$2.249\times10^{6}$  & $4.467\times10^{6}$ 
\rule{0pt}{10pt} \\ 
	$^{230}_{94}$Pu& 7.178 & $1.050\times10^{2}$ &0.593& 0.526&
$9.050\times10^{2}$&	$3.582\times10^{2}$ & $1.731\times10^{2}$ 
 	& -- & $2.455\times10^{2}$ 
\rule{0pt}{10pt} \\ 
	$^{232}_{90}$Th& 4.082 & $4.418\times10^{17}$ &0.574& 0.510&
$1.816\times10^{18}$&	$3.569\times10^{17}$ & $1.644\times10^{17}$  
	&$8.670\times10^{17}$ & $1.259\times10^{18}$  
\rule{0pt}{10pt} \\ 
	$^{232}_{92}$U& 5.414 & $2.174\times10^{9}$ &0.582& 0.517&
$1.404\times10^{10}$&	$2.993\times10^{9}$ & $1.399\times10^{9}$ 
	&$2.979\times10^{9}$  & $4.571\times10^{9}$ 
\rule{0pt}{10pt} \\ 
	$^{234}_{92}$U& 4.858 & $7.747\times10^{12}$ &0.577& 0.512&
$3.731\times10^{13}$&	$4.938\times10^{12}$ & $2.272\times10^{12}$ 
	&$9.954\times10^{12}$  & $1.175\times10^{13}$ 
\rule{0pt}{10pt} \\ 
	$^{236}_{92}$U& 4.573 & $7.391\times10^{14}$ &0.575& 0.511&
$3.742\times10^{15}$&	$4.810\times10^{14}$ & $2.196\times10^{14}$
	&$1.205\times10^{15}$  & $1.479\times10^{15}$ 
\rule{0pt}{10pt} \\ 
	$^{236}_{94}$Pu& 5.867 & $9.019\times10^{7}$ &0.583& 0.517&
$5.215\times10^{8}$&	$6.057\times10^{7}$ & $2.797\times10^{7}$ 
	&$8.630\times10^{7}$  & $9.120\times10^{7}$ 
\rule{0pt}{10pt} \\
	$^{238}_{92}$U& 4.270 & $1.408\times10^{17}$ &0.573& 0.509&
$8.205\times10^{17}$&	$8.777\times10^{16}$ & $3.978\times10^{16}$ 
	&$3.273\times10^{17}$  & $3.981\times10^{17}$ 
\rule{0pt}{10pt} \\
	$^{238}_{94}$Pu& 5.593 & $2.768\times10^{9}$ &0.581& 0.516&
$1.473\times10^{10}$&	$1.700\times10^{9}$ & $7.843\times10^{8}$ 
	&$2.723\times10^{9}$  &  $3.020\times10^{9}$ 
\rule{0pt}{10pt} \\ 
	$^{240}_{94}$Pu& 5.256 & $2.070\times10^{11}$ &0.579& 0.513&
$1.273\times10^{12}$&	$1.211\times10^{11}$ & $5.512\times10^{10}$ 
	&$2.786\times10^{11}$ & $2.754\times10^{11}$  
\rule{0pt}{10pt} \\
	$^{240}_{96}$Cm& 6.398 & $2.627\times10^{6}$ &0.584& 0.517&
$1.258\times10^{7}$&	$1.509\times10^{6}$ & $7.018\times10^{5}$
 & $1.637\times10^{6}$  	& $1.905\times10^{6}$
\rule{0pt}{10pt} \\
	$^{240}_{98}$Cf& 7.711 & $4.091\times10^{1}$ &0.591& 0.523&
$5.150\times10^{2}$&	$6.840\times10^{1}$ & $3.252\times10^{1}$ 
& $4.508\times10^{1}$  	& $5.623\times10^{1}$ 
\rule{0pt}{10pt} \\ 	 
	$^{242}_{94}$Pu& 4.984 & $1.183\times10^{13}$ &0.577& 0.512&
$6.506\times10^{13}$&	$6.092\times10^{12}$ & $2.750\times10^{12}$ 
	 & $1.667\times10^{13}$ & $1.778\times10^{13}$ 
\rule{0pt}{10pt} \\
	$^{242}_{96}$Cm& 6.216 & $1.407\times10^{7}$ &0.584& 0.518&
$8.497\times10^{7}$&	$1.058\times10^{7}$ & $4.851\times10^{6}$ 
	& $1.225\times10^{7}$ 	& $1.514\times10^{7}$ 
\rule{0pt}{10pt} \\  
	$^{244}_{94}$Pu& 4.666 & $2.569\times10^{15}$ &0.574& 0.510&
$1.019\times10^{16}$&	$1.083\times10^{15}$ & $4.899\times10^{14}$ 
	 & $3.236\times10^{15}$ & $3.890\times10^{15}$ 
\rule{0pt}{10pt} \\ 	 
	$^{244}_{96}$Cm& 5.902 & $5.715\times10^{8}$ &0.582& 0.516&
$3.060\times10^{9}$&	$3.667\times10^{8}$ & $1.668\times10^{8}$ 
	& $4.966\times10^{8}$  & $6.607\times10^{8}$ 
\rule{0pt}{10pt} \\ 
	$^{244}_{98}$Cf& 7.329 & $1.560\times10^{3}$ &0.591& 0.524&
$1.179\times10^{4}$&	$1.591\times10^{3}$ & $7.463\times10^{2} $
	&$1.250\times10^{3}$  & $1.660\times10^{3}$ 
\rule{0pt}{10pt} \\ 
	$^{246}_{96}$Cm& 5.475 & $1.485\times10^{11}$ &0.578& 0.513&
$6.820\times10^{11}$&	$9.222\times10^{10}$ & $4.173\times10^{10}$ 
	& $1.429\times10^{11}$  & $2.042\times10^{11}$ 
\rule{0pt}{10pt} \\ 
	$^{246}_{98}$Cf& 6.862 & $1.285\times10^{5}$ &0.587& 0.520&
$8.799\times10^{5}$&	$1.093\times10^{5}$ & $5.042\times10^{4}$ 
	& $9.954\times10^{4}$  & $1.318\times10^{5}$ 
\rule{0pt}{10pt} \\ 
	$^{246}_{100}$Fm& 8.379 & $1.652\times10^{0}$&0.597& 0.529&
$1.737\times10^{1}$&	$2.973\times10^{0}$ & $1.410\times10^{0}$
	&$1.592\times10^{0}$  & $2.344\times10^{0}$ 
\rule{0pt}{10pt} \\ 
	$^{248}_{96}$Cm& 5.162 & $1.199\times10^{13}$ &0.576& 0.510&
$5.521\times10^{13}$&	$8.361\times10^{12}$ & $3.751\times10^{12}$ 
	& $1.216\times10^{13}$  	& $2.188\times10^{13}$ 
\rule{0pt}{10pt} \\  
	$^{248}_{98}$Cf& 6.361 & $2.881\times10^{7}$ &0.582& 0.516&
$1.504\times10^{8}$&	$2.141\times10^{7}$ & $9.702\times10^{6}$ 
	& $1.919\times10^{7}$  & $3.388\times10^{7}$ 
\rule{0pt}{10pt} \\ 
	$^{248}_{100}$Fm& 7.995 & $3.450\times10^{1}$&0.595& 0.527&
$3.145\times10^{2}$&	$5.128\times10^{1}$ & $2.391\times10^{1}$ 
	&$2.679\times10^{1}$  & $4.677\times10^{1}$ 
\rule{0pt}{10pt} \\ 
	$^{250}_{98}$Cf& 6.128 & $4.131\times10^{8}$ &0.581& 0.515&
$1.989\times10^{9}$&	$3.176\times10^{8}$ & $1.434\times10^{8}$ 
	& $2.793\times10^{8}$  & $5.495\times10^{8}$ 
\rule{0pt}{10pt} \\
	$^{250}_{100}$Fm& 7.557 & $1.860\times10^{3}$&0.591& 0.523&
$1.127\times10^{4}$&	$1.999\times10^{3}$ & $9.238\times10^{2}$
	&$1.064\times10^{3}$  & $2.089\times10^{3}$ 
\rule{0pt}{10pt} \\   
	$^{252}_{98}$Cf& 6.217 & $8.614\times10^{7}$ &0.570& 0.507&
$5.093\times10^{8}$&	$9.631\times10^{7}$ & $4.559\times10^{7}$ 
	& $9.817\times10^{7}$  & $2.089\times10^{8}$ 
\rule{0pt}{10pt} \\   
	$^{252}_{100}$Fm& 7.154 & $9.140\times10^{4}$&0.588& 0.520&
$4.025\times10^{5}$&	$7.785\times10^{4}$ & $3.584\times10^{4}$ 
	&$4.064\times10^{4}$  & $9.120\times10^{4}$ 
\rule{0pt}{10pt} \\ 
\bottomrule
\end{longtable*}

\begin{longtable*}{@{\extracolsep{\fill}} l l l l l l l l l l @{}} 
	\caption{Same as Table \ref{tb4}, but for cluster decay half-lives. \(G\) and \(\ell\) denote the global quantum number and the orbital angular momentum, respectively (see Eqs.~\eqref{eq1}, \eqref{eq13}, and \eqref{eq14}).} 
	\label{tb5} \\ \toprule
	Decay & $Q$ \cite{Wang21} 
	& $T_{1/2}^{\rm exp}$ \cite{Kondev21} 
    & $G$
    & $\ell$
    & $a^{\text{WS}}$
    & $a^{\text{mWS}}$
	& $T_{1/2}^\text{WS}\text{(Def)}$ 
	& $T_{1/2}^\text{mWS}\text{(Def)}$
	& $T_{1/2}$ \cite{Dumi22} \\
	process & (MeV)   & (s)
	&     
	&
    & $\text{(fm)}$  
    & $\text{(fm)}$  
	& (s) & (s)   & (s) \\  \midrule 
	\endfirsthead
	\multicolumn{7}{c}
	{\tablename\ \thetable\ -- \textit{Continued}} \\ [5pt]
	\hline
	Decay & $Q$ \cite{Wang21} 
	& $T_{1/2}^\text{exp}$ \cite{Kondev21}   
	& $G$
	& $\ell$
    & $a^{\text{WS}}$
    & $a^{\text{mWS}}$
	& $T_{1/2}^\text{WS}\text{(Def)}$ 
	& $T_{1/2}^\text{mWS}\text{(Def)}$ 
	& $T_{1/2}$ \cite{Dumi22} \\
	process & (MeV)        & (s)
	&  
	&
    & $\text{(fm)}$ 
    & $\text{(fm)}$    
	& (s) & (s)        & (s)  \\  \midrule 
	\endhead
	\multicolumn{8}{r}{\textit{Continued on next page}} \\
	\endfoot
	\endlastfoot
	$^{222}_{88}\mathrm{Ra} \rightarrow {}^{208}_{82}\mathrm{Pb} + {}^{14}_{6}\mathrm{C}$ & 33.049 &  $1.120\times10^{11}$ & 68 & 0 & 0.914 & 0.813
	& $1.672\times10^{11}$
	& $2.562\times10^{10}$
	& $2.773\times10^{12}$
\rule{0pt}{10pt} \\
\textcolor{black}{
$^{223}_{88}\mathrm{Ra} \rightarrow {}^{209}_{82}\mathrm{Pb} + {}^{14}_{6}\mathrm{C}$} & 31.828 &  $1.110\times10^{15}$ & 68 & 4 & 0.916 & 0.814
& $2.109\times10^{15}$
& $2.876\times10^{14}$
& --
\rule{0pt}{10pt} \\
	$^{224}_{88}\mathrm{Ra} \rightarrow {}^{210}_{82}\mathrm{Pb} + {}^{14}_{6}\mathrm{C}$ & 30.534 &  $7.844\times10^{15}$ & 68 & 0 & 0.913 & 0.812
	& $1.190\times10^{16}$
	& $1.464\times10^{15}$
	& $7.482\times10^{16}$
\rule{0pt}{10pt} \\
	$^{226}_{88}\mathrm{Ra} \rightarrow {}^{212}_{82}\mathrm{Pb} + {}^{14}_{6}\mathrm{C}$ & 28.197 &  $1.942\times10^{21}$ & 68 & 0 & 0.913 & 0.813
	& $3.427\times10^{21}$
	& $3.943\times10^{20}$
	& $4.436\times10^{21}$
\rule{0pt}{10pt} \\
\textcolor{black}{
$^{225}_{89}\mathrm{Ac} \rightarrow {}^{211}_{83}\mathrm{Bi} + {}^{14}_{6}\mathrm{C}$ } & 30.476 &  $1.617\times10^{17}$ & 68 & 4 & 0.916 & 0.814
& $3.420\times10^{17}$
& $4.155\times10^{16}$
&  --	
\rule{0pt}{10pt} \\
	$^{228}_{90}\mathrm{Th} \rightarrow {}^{208}_{82}\mathrm{Pb} + {}^{20}_{8}\mathrm{O}$ & 44.723 &  $5.341\times10^{20}$ & 92 & 0 & 0.963 & 0.856
	& $2.302\times10^{21}$
	& $1.809\times10^{20}$
	& $4.426\times10^{20}$
\rule{0pt}{10pt} \\
	$^{230}_{90}\mathrm{Th} \rightarrow {}^{206}_{80}\mathrm{Hg} + {}^{24}_{10}\mathrm{Ne}$ & 57.760 & $4.102\times10^{24}$ & 106 & 0 & 0.997 & 0.885
	& $4.163\times10^{25}$
	& $2.282\times10^{24}$
	& $5.649\times10^{23}$
\rule{0pt}{10pt} \\
	$^{232}_{92}\mathrm{U} \rightarrow {}^{208}_{82}\mathrm{Pb} + {}^{24}_{10}\mathrm{Ne}$ & 62.309 &  $2.443\times10^{20}$ & 106 & 0 & 1.001 & 0.887
	& $1.472\times10^{21}$
	& $9.352\times10^{19}$
	& $1.476\times10^{20}$
\rule{0pt}{10pt} \\
\textcolor{black}{
$^{233}_{92}\mathrm{U} \rightarrow {}^{209}_{82}\mathrm{Pb} + {}^{24}_{10}\mathrm{Ne}$ } & 60.485 &  $6.977\times10^{24}$ & 106 & 2 & 1.004 & 0.889
& $7.116\times10^{25}$
& $3.779\times10^{24}$
& --
\rule{0pt}{10pt} \\
	$^{234}_{92}\mathrm{U} \rightarrow {}^{210}_{82}\mathrm{Pb} + {}^{24}_{10}\mathrm{Ne}$ & 58.825 &  $8.608\times10^{25}$ & 106 & 0 & 1.001 & 0.887
	& $1.217\times10^{27}$
	& $6.003\times10^{25}$
	& $3.793\times10^{24}$
\rule{0pt}{10pt} \\
	$^{234}_{92}\mathrm{U} \rightarrow {}^{208}_{82}\mathrm{Pb} + {}^{26}_{10}\mathrm{Ne}$ & 59.413 &  $8.608\times10^{25}$ & 114 & 0 & 1.006 & 0.893
	& $1.103\times10^{27}$
	& $5.554\times10^{25}$
	& $9.705\times10^{25}$
\rule{0pt}{10pt} \\
	$^{234}_{92}\mathrm{U} \rightarrow {}^{206}_{80}\mathrm{Hg} + {}^{28}_{12}\mathrm{Mg}$ & 74.111 &  $5.534\times10^{25}$ & 120 & 0 & 1.028 & 0.910
	& $5.646\times10^{26}$
	& $2.299\times10^{25}$
	& $4.140\times10^{24}$
\rule{0pt}{10pt} \\
	$^{236}_{94}\mathrm{Pu} \rightarrow {}^{208}_{82}\mathrm{Pb} + {}^{28}_{12}\mathrm{Mg}$ & 79.670 & $4.509\times10^{21}$ & 120 & 0 & 1.032 & 0.912
	& $2.537\times10^{22}$
	& $1.320\times10^{21}$
	& $1.050\times10^{21}$
\rule{0pt}{10pt} \\
	$^{238}_{94}\mathrm{Pu} \rightarrow {}^{210}_{82}\mathrm{Pb} + {}^{28}_{12}\mathrm{Mg}$ & 75.911 & $4.613\times10^{25}$ & 120 & 0 & 1.031 & 0.912
	& $1.083\times10^{27}$
	& $4.268\times10^{25}$
	& $9.354\times10^{24}$
\rule{0pt}{10pt} \\
	$^{238}_{94}\mathrm{Pu} \rightarrow {}^{208}_{82}\mathrm{Pb} + {}^{30}_{12}\mathrm{Mg}$ & 76.793 &  $4.613\times10^{25}$ & 128 & 0 & 1.035 & 0.917
	& $8.537\times10^{26}$
	& $3.559\times10^{25}$
	& $6.776\times10^{25}$
\rule{0pt}{10pt} \\
	$^{238}_{94}\mathrm{Pu} \rightarrow {}^{206}_{80}\mathrm{Hg} + {}^{32}_{14}\mathrm{Si}$ & 91.187 &  $1.977\times10^{25}$ & 134 & 0 & 1.053 & 0.931
	& $3.627\times10^{26}$ 
	& $1.344\times10^{25}$
	& $1.589\times10^{25}$
\rule{0pt}{10pt} \\
	$^{242}_{96}\mathrm{Cm} \rightarrow {}^{208}_{82}\mathrm{Pb} + {}^{34}_{14}\mathrm{Si}$ & 96.544 &  $1.279\times10^{23}$ & 142 & 0 & 1.061 & 0.938
	& $1.289\times10^{24}$
	& $6.243\times10^{22}$
	& $1.560\times10^{24}$
\rule{0pt}{10pt} \\
	\bottomrule 
\end{longtable*}


%

\end{document}